\documentclass[a4paper,fleqn]{cas-sc}
\usepackage[authoryear]{natbib}

\usepackage{url} 
\usepackage{lineno}
\usepackage{empheq}
\usepackage{amsmath,amsfonts}
\usepackage{enumitem}
\usepackage{subcaption}
\usepackage{booktabs}
\usepackage{algorithm}
\usepackage[noend]{algpseudocode}
\renewcommand{\d}{\mathrm{d}}
\newcommand{\f}{\mathrm{f}}
\newcommand{\Ki}{\mathrm{k}}
\newcommand{\Kh}{K}
\newcommand{\m}{\textrm{m}}
\newcommand{\n}{\textrm{n}}
\newcommand{\Flux}{\textbf{q}}
\newcommand*\widefbox[1]{\fbox{\hspace{2em}#1\hspace{2em}}}

\def\tsc#1{\csdef{#1}{\textsc{\lowercase{#1}}\xspace}}
\tsc{WGM}
\tsc{QE}
\tsc{EP}
\tsc{PMS}
\tsc{BEC}
\tsc{DE}

\begin{document}
\let\WriteBookmarks\relax
\def\floatpagepagefraction{1}
\def\textpagefraction{.001}
\shorttitle{A model and numerical method for gravity-driven variably saturated groundwater flow}

\shortauthors{Shadab and Hesse}

\title [mode = title]{A hyperbolic-elliptic PDE model and conservative numerical method for gravity-dominated variably-saturated groundwater flow}
\tnotemark[1,2]

\tnotetext[1]{This research project was supported by National Aeronautics and Space Administration under Emerging World Grant numbers NASA $18-$EW$18\_ 2-0027$ and \#80NSSC19K0505 of M.A.H. and University of Texas Institute for Geophysics under Blue Sky Student Fellowship of M.A.S. The authors acknowledge the discussions with Cyril Grima and Anja Rutishauser that motivated this work.}
\tnotetext[2]{All figures are colored. Code is available on GitHub: \url{https://github.com/mashadab/VarSatFlow}}
%
\author[1,2]{Mohammad Afzal Shadab}[type=editor,
                        orcid=0000-0002-0797-5017]

\cormark[1]


\ead{mashadab@utexas.edu}

\ead[url]{https://mashadab.github.io/}

\credit{Conceptualization of this study, Methodology, Software, Data curation, Writing - Original draft preparation}

\address[1]{Oden Institute for Computational Engineering and Sciences, The University of Texas at Austin, Austin TX 78712, United States}


\author[1,3]{Marc Andre Hesse}[orcid=0000-0002-2532-3274
   ]
                       
\ead{mhesse@jsg.utexas.edu}
\ead[URL]{https://www.jsg.utexas.edu/hesse/marc-hesse/}

\credit{Conceptualization of this study, Methodology, Writing - Original draft preparation, Supervising}

\address[2]{University of Texas Institute for Geophysics, Austin TX 78758, United States}
    

\address[3]{Department of Geological Studies, Jackson School of Geosciences, The University of Texas at Austin, Austin TX 78712, United States}


\cortext[cor1]{Corresponding author}



\begin{abstract}
Richards equation is often used to represent two-phase fluid flow in an unsaturated porous medium when one phase is much heavier and more viscous than the other. {However, it cannot describe the fully saturated flow for some capillary functions without specialized treatment due to degeneracy in the capillary pressure term.} Mathematically, gravity-dominated variably saturated flows are interesting because their governing partial differential equation switches from hyperbolic in the unsaturated region to elliptic in the saturated region. Moreover, the presence of wetting fronts introduces strong spatial gradients often leading to numerical instability. {In this work, we develop a robust, multidimensional mathematical model and implement a well-known efficient and conservative numerical method for such variably saturated flow in the limit of negligible capillary forces.} The elliptic problem in saturated regions is integrated efficiently into our framework by solving a reduced system corresponding only to the saturated cells using fixed head boundary conditions in the unsaturated cells. In summary, this coupled hyperbolic-elliptic PDE framework provides an efficient, physics-based extension of the hyperbolic Richards equation to simulate fully saturated regions. Finally, we provide a suite of easy-to-implement yet challenging benchmark test problems involving saturated flows in one and two dimensions. These simple problems, accompanied by their corresponding analytical solutions, can prove to be pivotal for the code verification, model validation (V\&V) and performance comparison of simulators for variably saturated flow. Our numerical solutions show an excellent comparison with the analytical results for the proposed problems. The last test problem on two-dimensional infiltration in a stratified, heterogeneous soil shows the formation and evolution of multiple disconnected saturated regions.

\end{abstract}




\maketitle

\section{Introduction}

Richardson-Richards equation \citep{richardson1922weather,Richards1931}, popularly known as Richards equation, describes the flow of water in an unsaturated porous medium due to gravity and capillary forces. It plays a crucial role in soil hydrology, agriculture, environment and waste management \citep{Farthing2017}. More recently, it has been implemented to study the meltwater percolation in glacier firn (porous, sintered and compacted snow) \citep{Colbeck1972,Meyer2017}. {Richards equation can be derived from the standard formulation of two-phase flow in porous media when there is a large contrast in the viscosity and density (mobility) of the two phases \citep{szymkiewicz2013mathematical,lie2019introduction}}. The large contrast in mobility leads to very high speeds of lighter, less viscous gas phase compared to heavier, more viscous water phase {\citep{Barenblatt1984}}. Thus, in the full two-phase flow model for a gas-water system the fast gas fronts become very restrictive for an explicit time integration due to the CFL condition. To avoid this, the contrast in mobilities is utilized for model simplification by neglecting the motion of the lighter phase via omitting the pressure gradient required to drive it. This results in Richards equation that captures the motion of the heavier, more viscous phase (water). 

Mathematically Richards equation is a nonlinear, parabolic partial differential equation. It transitions from parabolic to degenerate elliptic as the (sub-)domain nears complete saturation \citep{list2016study}. More specifically, the solution depends on two highly nonlinear soil water constitutive functions which depend on water saturation ($s_w$), the hydraulic conductivity, $K(s_w)$, and the capillary suction head, $\Psi(s_w)$, where the latter approaches zero rapidly in the near saturation limit ($s_w \to 1-s_{gr}$ with $s_{gr}$ being the residual gas saturation). Therefore it leads to an unbounded capillary suction head derivative term when the medium becomes saturated leading to a degeneracy, i.e., $|{\d \Psi}/{\d s_w}|\to \infty$ as $s_w \to 1-s_{gr}$ \citep{Farthing2017}. 
Furthermore, the infiltration into dry soils {or simulations with large spatial scales} often leads to the formation of wetting fronts (shock waves) causing extremely sharp gradients of soil hydraulic properties such as hydraulic conductivity and capillary suction head, leading to instability of the numerical solvers {\citep{Farthing2017,zha2019review}}. {At smaller spatial scales,} the term involving second-order derivative of capillary suction leads to diffusion of the wetting front and thus helps stabilize the numerical model. In summary, these nonlinearities and the degeneracy make the design and analysis of numerical schemes for the Richards equation very difficult {\citep{miller2013numerical}}. 

Nonetheless{,} Richards equation has been extensively used in the field of hydrology \citep{touma1986experimental,Farthing2017}, because variably saturated flows are the crucial link between surface water and groundwater. There are primarily three approaches to overcome the difficulties associated with simulating variably saturated flows using Richards equation \citep{zha2019review}. The first is to use the head-based form of Richards equation where the dependent variable is the hydraulic head, $h$, so that the derivative of capillary suction head, $\Psi$, with respect to saturation can be avoided {\citep{gillham1976hydraulic,Celia1990,zadeh2011mass,Farthing2017,zha2019review}}. {But the computationally-expensive head-based forms introduce mass balance errors in the numerical model \citep{Celia1990} and are more susceptible to numerical instability \citep{zha2019review}}. The second alternative is to modify the water retention curve so that its slope in the limit of complete saturation remains finite, i.e., $\d \Psi / \d s_w > - \infty$ when $s_w \to 1-s_{gr}$ {\citep{Clapp1978,vogel2000effect,keita2021implicit}}. The third approach is called the primary variable switching technique {\citep{zeng2018switching}} where the primary variable in the Richards equation is switched from water saturation to head depending on the cell saturation. However, this non-smooth transition between the two primary variables potentially produces unphysical solutions \citep{zha2019review}. Also, these switching techniques \citep{zadeh2011mass,zeng2018switching} are restricted to the expensive Newton-Raphson iterations \citep{zha2019review}. Additionally, {\cite{he2015comments}} only utilizes the mixed form in one dimension as the time derivative of the saturation term in Richards' equation numerically converges to zero at complete saturation. In summary, the solution of Richards equation remains challenging, in particular in variably saturated flows and with strong capillary gradients.

Motivated by large-scale applications and gravity-dominated flows, where these challenges are particularly prominent, we introduce a different approach to model variably saturated flow. In these situations, the flow is driven by gravity and the capillary gradients are very sharp so that they can not be resolved in field-scale applications. We therefore consider Richards equation in the limit of negligible capillary forces to obtain a description of wetting fronts and the groundwater table. This leads to a mathematically interesting set of governing equations that are hyperbolic in the unsaturated region and elliptic in the saturated region. Recently, similar approaches for treating the saturated region have been developed by \cite{Meyer2017} and {\cite{dai2019modeling}} for the full Richards equation with capillary effects. \cite{dai2019modeling} have shown their approach to be more robust than Richards equation based Hydrus-1D simulations {\citep{vsimunek2012hydrus}}, under conditions where a saturated region forms. However, both approaches are limited to one-dimension, where the saturated region can be integrated analytically. \cite{shadab2022analysis} {developed a model for gravity dominated flow with the formation of saturated regions in the limit of no capillary forces and provide analytic solutions for} one-dimensional {infiltration} problems. {Here we extend their model to multiple dimensions and develop a conservative numerical method that naturally captures the interactions of multiple saturated domains.}

In addition to the models, ``there is also a need of benchmark test problems to facilitate consistent advances and avoid reinventing of the wheel'', according to \cite{Farthing2017}. There are several specialized benchmark problems proposed in the integrated hydrologic model comparison project {\citep{kollet2017integrated}} and by the International Soil Modeling Consortium (\url{https://soil-modeling.org/}) {\citep{vereecken2016modeling,baatz2019international}}. These benchmarks consider complex scenarios aimed at testing fully developed simulators. Here we offer an additional set of relatively simple tests that focus solely on the interaction of saturated and unsaturated regions. All cases have semi-analytic solutions and/or allow direct comparison with experimental data for code verification and model validation.

{It is noted that extensive theoretical and numerical studies in the field of emerging computational geosciences \citep{zhao2009fundamentals} have demonstrated that physical and chemical dissolution reactions in porous media can cause variations of porosity, so that groundwater flow is fully coupled with physical and chemical dissolution reactions, porosity evolution and mass transport processes \citep{zhao2008effect,zhao2010theoretical}. In particular, the process of a mineral dissolution reaction is affected by the following factors such as the mineral dissolution ratio \citep{zhao2010effects}, mineral surface shape \cite{zhao2008effect}, solute dispersion \citep{zhao2010theoretical}, porous medium compressibility \citep{zhao2014effects}, porous medium anisotropy \citep{zhao2013effects}, and non-isothermal influence \citep{Colbeck1972,shadab2024unified,shadab202Xicelayer,zhao2015theoretical}. More importantly, due to the full coupling of groundwater flow with the above-mentioned three main processes, the porosity and related physical parameters are no longer constant in realistic flow systems \cite{zhao2013theoretical}. Therefore, physical and chemical dissolution reactions should be considered to reflect this reality \citep{zhao2017new,zhao2020transient}. However, below we focus on the variably saturated flow in the limit of negligible capillary forces.}


In this paper we propose a multidimensional, physics-based extension to the Richards model to solve variably saturated flow while neglecting  capillary forces. 
Section \ref{sec2:hypwatertransporteq} introduces the theoretical framework that extends the hyperbolic Richards equation model for unsaturated flow to the case of complete saturation. Section \ref{sec3:solver} presents the numerical model along with the proposed hyperbolic-elliptic PDE solution algorithm. The proposed solver robustly and efficiently simulates variably saturated flow with sharp gradients and fully saturated elliptic region(s) where other Richards equation based simulators may fail. Section \ref{sec:test-problems} documents a suite of 1D and 2D benchmark test problems with known analytical solutions and the validation of our numerical model against it. Section \ref{sec5:conclusions} concludes the present work.

\section{Model formulation}\label{sec2:hypwatertransporteq}

\begin{figure}
    \centering
    \includegraphics[width=0.33\linewidth]{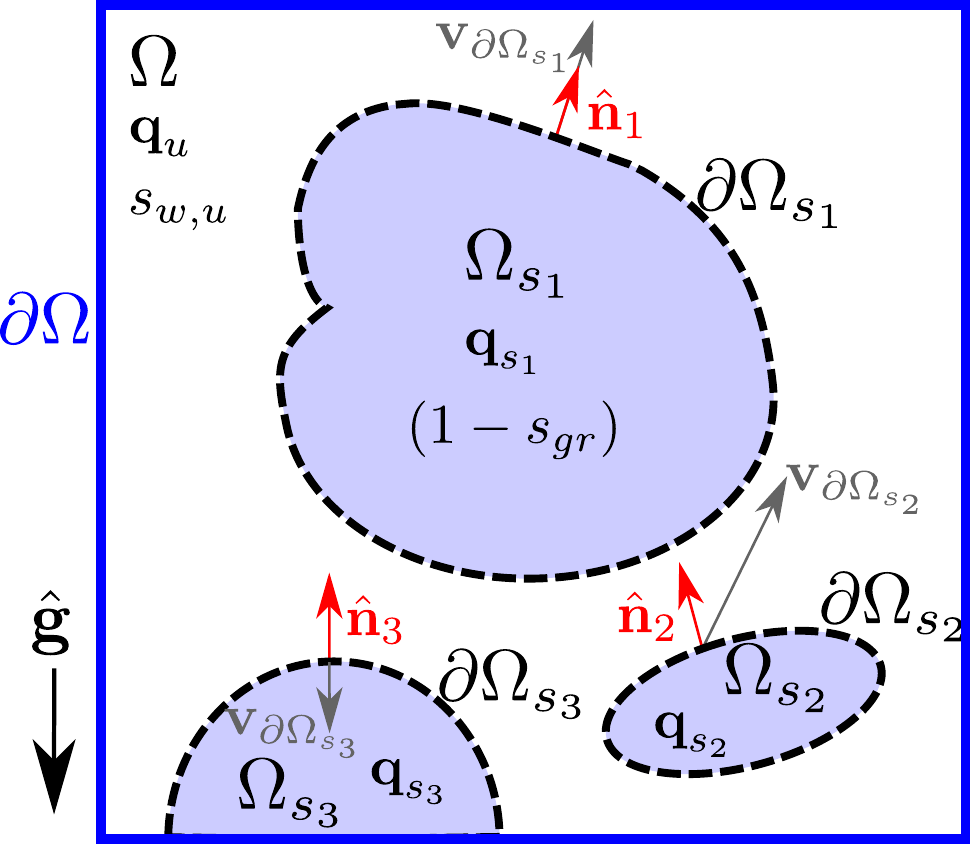}
    \caption{A schematic diagram illustrating an example of a variably saturated flow inside a domain $\Omega$ with boundary $\partial \Omega$ illustrated with solid blue line. The otherwise unsaturated domain (white region) has three fully saturated subdomains (shaded region) constituting $\Omega_s \equiv \Omega_{s_1} \cup \Omega_{s_2} \cup \Omega_{s_3}$. Also, for the $k^\textrm{th}$ saturated subdomain ($k\in\{1,2,3 \}$), $\hat{\textbf{n}}_k$ is the outward normal vector to the corresponding saturated-unsaturated region boundary $\partial \Omega_{s_k}$ and $\textbf{q}_{s_k}$ is the spatially varying volumetric flux of water phase, at a specific location.}
    \label{fig:potato_diagram}
\end{figure}

\subsection{Full two-phase fluid flow model}
Consider a system of two incompressible fluid phases, water and gas, in a non-deforming, stationary and porous medium with porosity $\phi$ (-) and permeability $\Ki$ (m$^\textrm{2}$). {Both of these fields ($\phi,k$) can thus vary in space but are time invariant.} The transport equations for these two phases can be written as

\begin{align}
\phi \frac{\partial}{\partial t } (\rho_w s_w) + \nabla \cdot (\rho_w \textbf{q}_w)  &= 0 \quad \forall \textbf{x} \in \Omega  \backslash \partial \Omega,~ t \in (0,T],  \label{eq:mb1}\\
\phi \frac{\partial }{\partial t }(\rho_g s_{g})+ \nabla \cdot (\rho_g \textbf{q}_g)  &=0  \quad \forall \textbf{x} \in \Omega  \backslash \partial \Omega,~ t \in (0,T],  \label{eq:mb2}
\end{align}
where subscripts $w$ and $g$ refer to the variables corresponding to water and gas phases respectively. The saturation (-), density (kg/m$^\text{3}$) and volumetric flux vector (m$^\text{3}$/(m$^\text{2}$.s)) of phase $\alpha \in \{ w,g\}$ are given by $s_\alpha$, $\rho_\alpha$ and $\textbf{q}_\alpha$, respectively. Moreover, $\Omega$ is the spatial domain with boundary $\partial \Omega$ {(see Figure \ref{fig:potato_diagram} for example)}, $\textbf{x}\equiv \left( x,y,z\right)^T$ is the spatial coordinate vector (m), $t$ is the time variable (s) and $T$ is the final time (s). {The backslash symbol ($\backslash$) denotes the set exclusion. For example, $A\backslash B$ means elements of set $A$ which are not present in set $B$.} 
Equations \eqref{eq:mb1} and \eqref{eq:mb2} can be combined to yield the two-phase continuity equation

\begin{eqnarray} \label{eq:mb3}
\nabla \cdot ( \textbf{q}_w + \textbf{q}_g) = 0  \quad \forall \textbf{x} \in \Omega.
\end{eqnarray}
The volumetric flux of each phase is given by an extension to Darcy's law as

\begin{equation} \label{eq:1darcy}
    \textbf{q}_{\alpha} = -\frac{ \Ki k_{r\alpha}}{\mu_\alpha} (\nabla p_\alpha - \rho_\alpha \textbf{g}), \quad \alpha \in \{w,g\},
\end{equation}
where for each fluid phase, $k_{r\alpha}$ is the relative permeability (-), $p_\alpha$ is the pressure (Pa) and $\mu_\alpha$ is the dynamic viscosity (Pa s). Relative permeabilities, $k_{r \alpha}$, are function of water saturation, $s_w$ \citep{Wyckoff1936}. The absolute permeability of the porous medium (m$^\text{2}$), given by $\Ki$, is a function of the porosity, $\phi$. The symbol $\textbf{g}$ represents the acceleration due to gravity vector (m/s$^\text{2}$) where $\textbf{g}=g \hat{\textbf{g}}$ with $\hat{\textbf{g}}$ being the unit vector in the direction of gravity {(see Figure \ref{fig:potato_diagram})}. 

The system of equations is closed by the constitutive relation for capillary pressure, $p_c$, as
\begin{align}
   p_c(s_w) = p_g-p_w,\label{eq:cap pres}
\end{align}
which relates the two phase pressures and is typically assumed to be a function of saturation only \citep{Leverett1941,brooks1964hydrau,VanGenuchten1980}. The constitutive functions for multi-phase flow, $k_{r \alpha}$ and $p_c$,  display complex hysteresis \citep{Blunt2017}, but here we only consider the simplest case where each phase becomes immobile below a certain residual saturation, $s_{\alpha r}$. So that the two-phase flow is restricted to regions where $ s_{wr}<s_w<1-s_{gr}$. In this paper we refer to regions with $s_w=1-s_{gr}$ as saturated, although immobile gas bubbles are still present. In the next section we transition to the Richards limit and build a connection with full two-phase flow formulation which will pave way for a physics-based model to simulate variably saturated flow with negligible capillary effects.

\subsection{{Modification of Richards equation for variably saturated flow with negligible capillary effects}}
For rainwater infiltration into soil both density and viscosity of gas are much smaller than that of water, i.e., $\rho_g \ll \rho_w$ and $\mu_g \ll \mu_w$.
In this limit, the gas responds essentially instantaneously and its pressure can be assumed a constant, so that only the flow of the water phase is considered. In this limit, the full two-phase flow equations (\ref{eq:mb1}-\ref{eq:cap pres}) reduce to Richards equation \citep{Richards1931}, as discussed in {\cite{szymkiewicz2013mathematical} and \cite{lie2019introduction}}. The saturation form of the Richards equation is given by

\begin{align}
\phi\frac{\partial s_w}{\partial t } + \nabla \cdot \left[ \Kh(s_w) \left( \nabla { \Psi(s_w)}+\hat{\textbf{g}}\right)\right]  = 0 \quad \forall \textbf{x} \in \Omega \backslash \partial \Omega , t \in (0,T], \label{eq:Richards}
\end{align}
where  $\Kh(s_w)=\Ki k_{rw}(s_w)\rho_w g/\mu_w$ is the saturation-dependent hydraulic conductivity (m/s), $\Psi(s_w) = p_c(s_w)/(\rho_w g)$ is the capillary suction head (m). 
The capillary effects can be neglected in the limit when $|\nabla \Psi| \ll 1$ \citep{Smith1983} which is a good approximation for flow in light textured soils such as sand {\citep{brustkern1970analytical,morel1974derivation,smith2002infiltration}} or for hydrological problems with large spatial scales {(see \cite{shadab2022analysis,Smith1983})}. {In \cite{shadab2022analysis} we present a detailed analysis of the effect of neglecting capillary suction in typical soil infiltration experiments.} This simplification leads to a gravity-driven unsaturated flow governed by

\begin{align}
  \phi \frac{\partial s_w}{\partial t } + \nabla \cdot \Kh(s_w) \hat{\textbf{g}}  = 0, \quad \forall \textbf{x} \in \Omega \backslash \partial \Omega , t \in (0,T]. \label{eq:hyperbolic-Richards}
\end{align}
This limit is commonly known as the kinematic wave approximation in the infiltration community \citep{Charbeneau1984,te2010applied,shadab2022analysis}. {However, for generality we refer to it as the hyperbolic Richards equation in this paper, due to the presence of only the gravity-driven advective flux term.} Equation \eqref{eq:hyperbolic-Richards} is naturally one-dimensional, if gravity is aligned with a coordinate direction. Even in a multi-dimensional and heterogeneous medium any unsaturated front, modeled through \eqref{eq:hyperbolic-Richards}, will migrate strictly in the direction of gravity, $\hat{\textbf{g}}$. This changes when the medium saturates locally and pressure gradients couple the flow in all directions across the saturated region. Since gas phase is immobile in a saturated region ($\textbf{q}_g=\textbf{0}$), Equations \eqref{eq:mb3} and \eqref{eq:1darcy} limit to the elliptic equation for incompressible saturated flow

\begin{align}\label{eqn:laplace}
    -\nabla\cdot\left(K\nabla h\right) &= 0  \quad \forall \textbf{x} \in \Omega_s(t) \textbackslash \partial\Omega_s(t),
\end{align}
where $h = p_w/(\rho_w g) - \int_0^\textbf{x}\hat{\textbf{g}} \cdot \d \textbf{x}$ is the hydraulic head (m). We have used $K$ to refer to the saturated hydraulic conductivity, which is strictly $K(1-s_{gr})$.
Solving variably saturated flow problems in the gravity-driven limit requires a dynamic coupling between the hyperbolic PDE \eqref{eq:hyperbolic-Richards} for unsaturated regions with elliptic PDE \eqref{eqn:laplace} for saturated regions. Although the elliptic PDE \eqref{eqn:laplace} itself is not time dependent, the saturated domain, $\Omega_s(t)$, changes with time due to its interaction with the unsaturated region. We refer to the interface between the saturated and unsaturated regions simply as the interface and denote it as $\partial\Omega_s(t)$ which may evolve with time. There can be multiple saturated regions that can dynamically form and evolve and can interact with each other (see Figure \ref{fig:potato_diagram}).

{Since the pressure in the unsaturated region is always determined by the gas phase and hence zero \citep{szymkiewicz2013mathematical,lie2019introduction,shadab2022analysis}}, the hydraulic head boundary condition along the interface is simply

\begin{align}\label{eq:pressure_bnd_cond}
    h=- \int_\textbf{0}^\textbf{x}\hat{\textbf{g}} \cdot \d \textbf{x} \quad \mathrm{on}\quad \mathbf{x}\in\partial\Omega_s(t),
\end{align}
where $\textbf{0}$ is the location vector of the origin, $(0,0,0)^T$. The {multidimensional} velocity of the interface, $\mathbf{v}_{\partial\Omega_s}$, can be determined by the discrete mass balance of water \eqref{eq:mb1} across the interface as

\begin{align}\label{eqn:vel_int}
    \mathbf{v}_{\partial\Omega_s} = \frac{(\mathbf{q}_u-\mathbf{q}_s)\cdot\hat{\mathbf{n}}}{\phi_u s_{w,u}-\phi_s(1-s_{gr})}\mathbf{\hat{n}},
\end{align}
where $\mathbf{\hat{n}}$ is the outward unit normal of the interface, $\mathbf{q}_u$ and $\mathbf{q}_s$ are the unsaturated and saturated fluxes along the interface {(see Figure \ref{fig:potato_diagram} for example)}. Additionally, $\phi_u s_{w,u}=\theta_u$ and $\phi_s(1-s_{gr})=\theta_s$ are the water volume fractions (soil moisture contents) on the unsaturated and saturated sides of the interface. The saturated domain, $\partial\Omega_s$, evolves according to this interface velocity. 

Due to the absence of the capillary term, which is singular in the saturated region, we can evolve the water saturation in both saturated and unsaturated domains. As such, we are
simply evolving the water mass balance \eqref{eq:mb1}, but evaluate the fluxes differently in the saturated and unsaturated regions. This avoids the explicit tracking of the interfaces and the mathematical model can be summarized as

\begin{empheq}[box=\widefbox]{align}
\phi \frac{\partial s_w}{\partial t }  + \nabla \cdot \textbf{q}_w  &= 0 \quad \forall \textbf{x} \in \Omega  \backslash \partial \Omega,~ t \in (0,T] \label{eq:final-sat-eqn}\\
\text{ with } \textbf{q}_w(s_w) &= \begin{cases}\textbf{q}_u=K(s_w)\hat{\textbf{g}} &\forall \textbf{x}\in \Omega \backslash \Omega_s \text{ where }  s_w < 1-s_{gr}  \\ \textbf{q}_s=-K \nabla h & \forall \textbf{x}\in \Omega_s \text{ where } s_w= 1-s_{gr}  \end{cases} \text{ and }  \label{eq:final-flux}\\
    -\nabla\cdot\left(K\nabla h\right) &= 0  \quad \forall \textbf{x} \in \Omega_s(t) \textbackslash \partial\Omega_s(t), \label{eqn:laplace-final} \\
    \textrm{subject to }       h&=- \int_\textbf{0}^\textbf{x}\hat{\textbf{g}} \cdot \d \textbf{x} \quad \forall \mathbf{x}\in\partial\Omega_s(t) \label{eq:pressure_bnd_cond-final}.
\end{empheq}

This hyperbolic-elliptic mathematical model (\ref{eq:final-sat-eqn}-\ref{eq:pressure_bnd_cond-final}) for variably-saturated flow in the limit of negligible capillary forces, requires the dynamic coupling of hyperbolic and elliptic subdomains with evolving interfaces. Below we develop a numerical algorithm that addresses the hyperbolic-elliptic nature of this model.

\section{Numerical model and algorithm} \label{sec3:solver}
\subsection{Discretization and operator approach}\label{sec:discrete-operator}
The governing equation is solved with conservative finite difference discretization on a regular Cartesian grid \citep{LeVeque1992} with total number of $N$ cells and $N_f$ faces. A staggered grid approach is used to avoid the checkerboard problem which leads to approximating the saturations and heads at the cell centers but fluid fluxes at cell faces \citep{ismail2010computational}. This formulation leads to second-order central differencing scheme in space. The tensor product grid enables a straightforward extension to multidimensions from one-dimension (1D). Here we briefly discuss the construction of two-dimensional (2D) operators from one-dimensional operators in the main text. See Appendix \ref{sec:one-D-operators} for the definition of one-dimensional operators and Appendix \ref{sec:3Dextension} for constructing operators in three-dimensions.

\textit{Gradient, divergence and mean operators}.  We use a regular Cartesian mesh with $n$ cells of size $\Delta x$ in $x$ direction and $m$ cells of size $\Delta z$ in $z$ direction. The two-dimensional discrete divergence operator, $\textbf{D} \in \mathbb{R}^{N\times N_f}$, discrete gradient operator, $\textbf{G} \in \mathbb{R}^{N_f \times N}$, and discrete mean operator, $\textbf{M} \in \mathbb{R}^{N_f \times N}$, are composed of two block matrices as

\begin{align} \label{eq:two-D-operators}
  \textbf{H} = \begin{bmatrix} \textbf{H}_x \\ \textbf{H}_z \end{bmatrix}_{N_f ~ \times~ N}\quad \mathrm{where}\quad \textbf{H}\in \{\textbf{D}^T~,~\textbf{G}~,~\textbf{M}\},
\end{align}
where $\mathbb{R}^k$ refers to the $k$ tuples of the real numbers. The matrices $\textbf{H}_x$ and $\textbf{H}_z$ are the $N_{fx}$ by $N$ and $N_{fz}$ by $N$ discrete operators in $x$ and $z$ directions respectively. In two-dimensions, $N_f$ is the total number of faces which is the summation of $x$ and $z$ normal faces, i.e., $N_f=N_{fx}+N_{fz}$ where $N_{fx}=(n+1)m$ and $N_{fz}=n(m+1)$. The total number of cells is $N=mn$. See Figure \ref{fig:HEalgo-schematic} as an example of a simple discretized domain with grid parameters given in the caption. Note that gradient and mean operators ($\textbf{G}$ and $\textbf{M}$ respectively) linearly map the cell centered variables to the corresponding cell faces whereas the divergence operator ($\textbf{D}$) does the opposite. Their corresponding block matrices, defined in Equation \eqref{eq:two-D-operators}, are obtained from the one-dimensional discrete operators using the Kronecker products

\begin{align} \label{eq:two-D-operators-sub}
    \textbf{H}_x = \textbf{H}_n \otimes \textbf{I}_m \quad & \text{and} \quad \textbf{H}_z = \textbf{I}_n \otimes \textbf{H}_m ,\quad \text{where the discrete operator} \quad \textbf{H}\in \{\textbf{D}~,~\textbf{G}~,~\textbf{M}\}
\end{align}
and $\textbf{I}_k$ is the $k$ by $k$ identity matrix. In this case, the sequence is chosen to respect the internal ordering of Matlab, where the cells and faces are ordered in $z$ direction first, then in $x$ direction. The one-dimensional discrete operators $\textbf{D}_\alpha$, $\textbf{G}_\alpha$ and $\textbf{M}_\alpha$ are given in Appendix~\ref{sec:one-D-operators} by Equations (\ref{eq:one-D-operators}-\ref{eq:one-D-mean}) with $\alpha =n$ and cell size $\Delta x$ for $x$ direction, and $\alpha =m$ and cell size $\Delta z$ for $z$ direction.

\textit{Advection operator}. For solving the saturation equation \eqref{eq:final-sat-eqn} the advective flux, given by $\textbf{q}_w$ in Equation \eqref{eq:final-flux}, is computed which requires the value of saturation (or soil moisture content) at the cell faces to evaluate the hydraulic conductivity. For that purpose, we use Darcy flux-based {first-order upwinding} to compute the saturation at the cell faces \citep{godunov1959finite}. In matrix form, we construct an advection matrix operator, $\textbf{A}(\textbf{q}_w) \in \mathbb{R}^{N_f \times N}$, which takes in the $N_f$ by $1$ Darcy flux vector, $\textbf{q}_w=[q_1,q_2,\hdots,q_{N_f-1},q_{N_f}]^T$, at the cell faces. The advection operator $\textbf{A}$ can be expressed in two dimensions as 

\begin{align}\label{eq:two-D-advection-operators1}
    \textbf{A} = \textbf{Q}^+\textbf{A}^+ ~+~ \textbf{Q}^-\textbf{A}^- \quad \text{with} \quad
\textbf{Q}^\pm =  \begin{bmatrix}
    \textbf{Q}_x^\pm & \\
    & \textbf{Q}_z^\pm
 \end{bmatrix}_{N_f~\times~N_f} \quad \text{and} \quad \textbf{A}^\pm =  \begin{bmatrix} \textbf{A}_x^\pm \\ \textbf{A}_z^\pm \end{bmatrix}_{N_f~\times~N}. 
 \end{align}
Here $\textbf{Q}_x^\pm$ is the $N_{fx}$ by $N_{fx}$ diagonal matrix with the positive or negative Darcy fluxes across the $x$ faces along the diagonal. Similarly $\textbf{Q}_z^\pm$ is the $N_{fz}$ by $N_{fz}$ diagonal matrix with $z$ face positive or negative Darcy fluxes in the diagonal. Moreover, the $x$ and $z$ matrix components of $\textbf{A}^+$ and $\textbf{A}^-$ are expressed similar to the 2D discrete operators discussed in Equation \eqref{eq:two-D-operators} as

\begin{align}\label{eq:two-D-advection-operators2}
     \textbf{A}_x^\pm = \textbf{A}_n^\pm \otimes \textbf{I}_m \quad & \text{and} \quad \textbf{A}_z^\pm = \textbf{I}_n \otimes \textbf{A}_m^\pm,
\end{align} where $\textbf{A}_n^\pm$ and $\textbf{A}_m^\pm$ are the one-dimensional operators, given by Equation \eqref{eq:one-D-advection-operators} in Appendix~\ref{sec:one-D-operators}.
In a similar fashion, this framework can be extended naturally to three-dimensions as described in Appendix~\ref{sec:3Dextension}.

\textit{Laplacian operator}. In this framework, the discrete Laplacian is simply given by $\textbf{L}=\textbf{D}\,\textbf{G}$, but here the discretization of the spatially variable hydraulic conductivity has to be considered. The hydraulic conductivity, ${K}(s_w)$, is approximated at cell centers and is therefore discretized as a $N$ by $1$ vector $\textit{\textbf{K}}$. The flux computation requires conductivities on the cell faces. Therefore $\textbf{\textit{K}}$ is harmonically averaged from cell centers to the faces. The $N_f$ by $1$ vector of averaged conductivities is given by $\textit{\textbf{K}}_m=(\textbf{M}\, \textit{\textbf{K}}^{-1})^{-1}$, where the algebraic mean operator, $\textbf{M}$, is used and the inverse refers to element wise reciprocal of the corresponding vector. The resulting vector $\textit{\textbf{K}}_m$ is stored as the $N_f$ by $N_f$ diagonal matrix \textbf{K}. The final spatial discretization of the heterogeneous Laplacian, $\textbf{L}$, in Equation \eqref{eqn:laplace-final} is then given by

\begin{align}
    -\nabla \cdot K(s_w) \nabla \approx \textbf{L}= -\textbf{D} ~ \textbf{K} ~ \textbf{G}.
\end{align}
So that for example, the elliptic equation \eqref{eqn:laplace-final} in the saturated region can be written in discrete form as

\begin{align}\label{eq:discrete-equation}
     -\nabla\cdot\left(K\nabla h\right) &= 0 \approx  \textbf{L}~ \textbf{h} = \textbf{f}_s
\end{align}
where $\textbf{h}$ and $\textbf{f}_s$ are $N$ by $1$ vectors of discrete unknown heads, $h(\mathbf{x})$, and known source terms, $f_s(\mathbf{x})$, at cell centers. For the problems considered in Section~\ref{sec:test-problems} there is no source term, so that $\textbf{f}_s = \textbf{0}$.

{The numerical method introduced in this section is discretely conservative in the way defined by \cite{LeVeque2002}, because the finite difference discretization of the conservative form of the governing equations is equivalent to a finite volume method on a tensor product mesh \citep{Aziz1979}. This ensures that jumps in saturation obey the Rankine-Hugoniot condition and that the discrete solution is a weak solution to Equation \eqref{eq:final-sat-eqn} \citep{Lax1960}.}

\subsection{Boundary conditions}\label{sec:BC-main}
Boundary conditions are required so that the PDE-based problem becomes well posed. {Natural boundary conditions} (zero gradient) are directly implemented in the 1D discrete gradient operator (Appendix~\ref{sec:one-D-operators}). Here we briefly summarize the implementation of other boundary conditions, for more details see Appendix \ref{sec:BC-full}. Non-homogeneous {Neumann boundary conditions} are applied by conversion of fluxes at the boundary into an equivalent extra source term in the corresponding boundary cells. For homogeneous {Dirichlet boundary conditions}, the number of unknowns is reduced in accordance with the $N_c$ number of constraints provided by the prescribed heads or saturations in the boundary cells. The constraints are eliminated by orthogonally projecting the solution vector, $\textbf{h}\in \mathbb{R}^N$, onto the null space of the constraints, $\mathcal{N} \in \mathbb{R}^{(N-N_c)}$ \citep{trefethen1997numerical}, and solving the resulting reduced system. For heterogeneous Dirichlet boundary conditions, the (quasi-)linearity of the problem allows splitting of the solution into homogeneous and particular solutions \citep{greenberg2013foundations}, followed by constraint elimination using the same orthogonal projection. 

\subsection{Time marching}
First-order, explicit Euler time integration is used to update the saturation equation \eqref{eq:final-sat-eqn}. {Since this temporal discretization along with first-order upwind scheme is conditionally stable \citep{hoffman2018numerical}, the CFL criteria is used.} The time step $\Delta t$ is evaluated from the minimum time among fastest filling of a cell and the CFL condition tracking the motion of fastest characteristic as

\begin{align} \label{eq:timestep}
    \Delta t = \min_{\Delta  t \in (0,\infty)}\left\{ \min_{\Delta  t \in (0,\infty),~\textbf{x}\in \Omega} \frac{|{\phi}(\textbf{x}) (1- s_{gr} - {s}_{w}(\textbf{x}))|}{\nabla \cdot \textbf{q}} , n_\mathrm{CFL} \times \frac{\min \{\Delta x, \Delta z \}}{ \max_{\textbf{x} \in \Omega}\lambda(\textbf{x)}} \right\}.
\end{align}
Here $\lambda$ is the speed of the characteristic at a location $\textbf{x}$ defined as $|{\d \textbf{q}_u(\textbf{x})}/{\d (\phi s_w)}|$ and $n_\mathrm{CFL}$ is the CFL number, $n_\mathrm{CFL}\in(0,1]$. The minimum time step evaluation requires a vectorized computation of Equation \eqref{eq:timestep} over the discretized domain. The first term inside curly braces in Equation \eqref{eq:timestep} corresponds to fastest fill time for cells along the saturated-unsaturated interface, which is typically the smallest time step when the domain contains a saturated region. Hence, the second term is redundant for the benchmark problems considered in Section~\ref{sec:test-problems}.

\subsection{Hyperbolic-elliptic PDE solver algorithm}
\begin{figure}
    \centering
    \includegraphics[width=0.95\linewidth,angle=90]{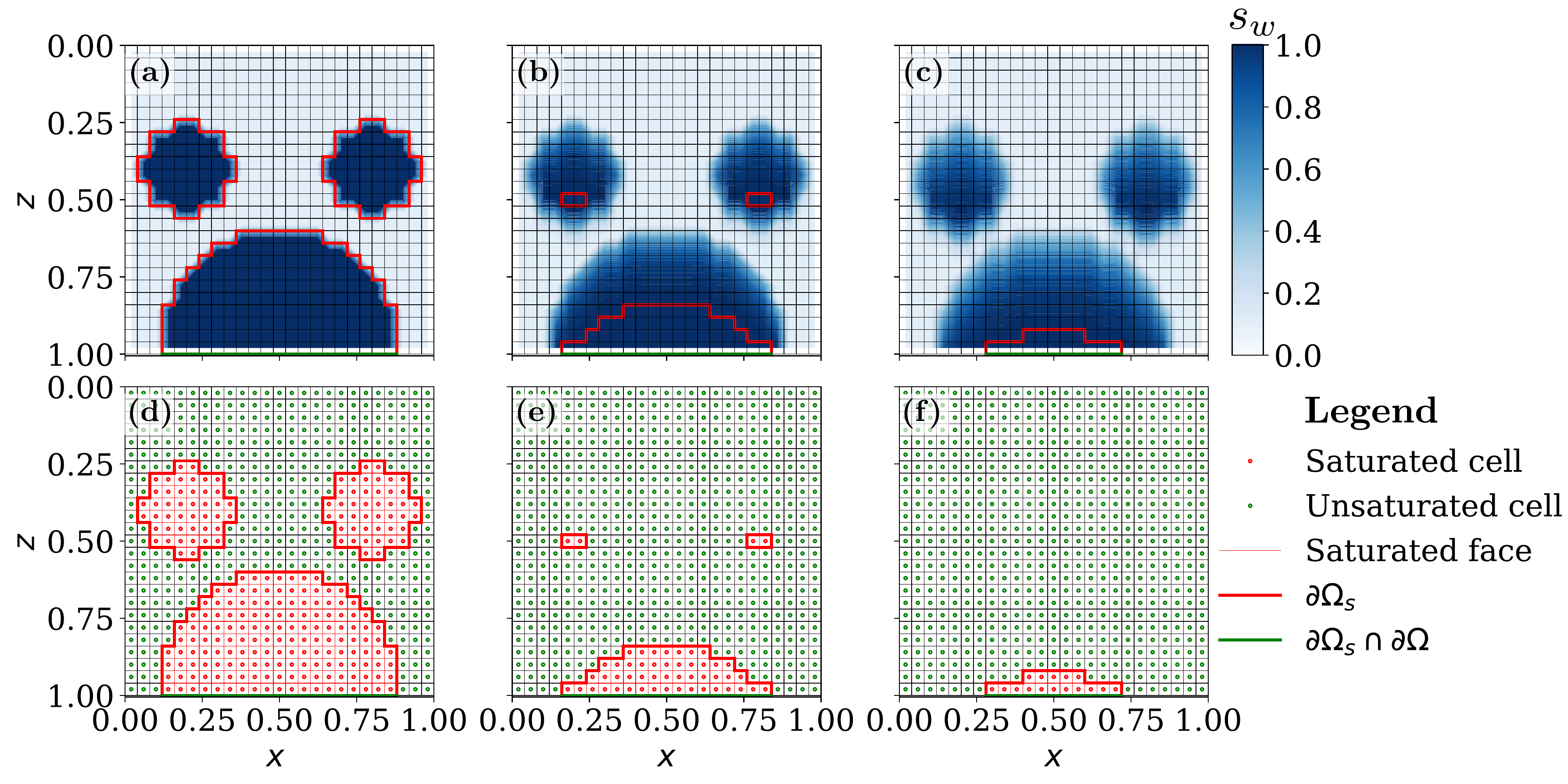}
    \caption{Gravity-dominated drainage of multiple saturated drops across an otherwise unsaturated ($s_w=10\% $) porous reservoir demonstrating the hyperbolic-elliptic PDE solution algorithm. The saturation contours are shown at dimensionless times $t'=$ (a) 0, (b) 0.025 and (c) 0.05. Their corresponding dynamically evolving saturated and unsaturated cells as well as faces are respectively shown in subfigures (d), (e) and (f). The gravity aligns with $+z$ direction, i.e., $\hat{\textbf{g}}=(0,0,1)^T$. The unit square domain is divided uniformly into $25 \times 25$ cells. Here the number of cells in $z$ and $x$ directions are $m=25$ and $n=25$ respectively and the total number of cells is $N=nm=25\times25=625$. The number of faces with normals in $z$ direction are $N_{fz}=n(m+1)=25\times26=650$ and in $x$ direction are $N_{fx}=(n+1)m=650$. The total number of cell faces are $N_f=N_{fx}+N_{fz}=1300$. }
    \label{fig:HEalgo-schematic}
\end{figure}

The proposed method aims to efficiently combine the simplicity of solving a single hyperbolic PDE for unsaturated flow with an additional, domain-specific elliptic PDE to capture the formation and evolution of fully-saturated regions (Figure~\ref{fig:HEalgo-schematic}\emph{a}-\ref{fig:HEalgo-schematic}\emph{c}). At the $i$-th time step the discretized saturation equation \eqref{eq:final-sat-eqn} takes the following form 

\begin{align}\label{eq:discrete-mass-balance}
    \boldsymbol{\Phi} \frac{\textbf{s}_w^{i+1} - \textbf{s}_w^i}{\Delta t^i} + \textbf{D}~\textbf{q}_w^i = \textbf{0},
\end{align}
where the superscripts denote the quantity at the corresponding time step. The symbol $\boldsymbol{\Phi}$ refers to $N$ by $N$ diagonal matrix with the known porosities of the cells as the diagonal entries, $\textbf{s}_w$ refers to the $N$ by $1$ vector of discretized water saturations at cell centers and $\textbf{q}_w$ is the $N_f$ by $1$ vector of the face-normal volumetric water fluxes evaluated at cell faces. This equation updates the water saturation vector, $\textbf{s}_w$, to the next time step ($i+1$). 

In the unsaturated region, governed by hyperbolic Richards equation \eqref{eq:hyperbolic-Richards}, the discretized flux $\textbf{q}_w^i$ can be simply evaluated at the cell faces by gravitational component of the Darcy flux \eqref{eq:final-flux} as $\textbf{q}_u=\textbf{K} ~\textbf{A}(\hat{\textbf{g}}) \textbf{k}_r(\textbf{s}_w^i)$ where $\textbf{k}_r$ is the $N$ by $1$ vector of normalized relative permeabilities, defined as $\textbf{k}_r=k_{rw}(\textbf{s}^i_w)/k_{rw}^0$, at cell centers. The saturated hydraulic conductivity matrix, $\textbf{K}$, is evaluated from the harmonic mean of the cell centered hydraulic conductivity vector, $\textit{\textbf{K}}$, whereas the $N$ by $1$ relative permeability vector, $\textbf{k}_r$, is upwinded in the direction of gravity using the advection operator, $\textbf{A}(\hat{\textbf{g}})$ (see Section \ref{sec:discrete-operator}). Here $\hat{\textbf{g}}$ is the $N_f$ by $1$ unit vector in the direction of gravity defined on the cell faces. If gravity is aligned with a coordinate direction then $\hat{\textbf{g}}$ is zero on faces with normals pointing in the other coordinate directions.  

Next the saturated subdomains comprising $\Omega_s$ are identified by selecting cells with saturations above a critical threshold, $s_{w,T}$. For example, Figures~\ref{fig:HEalgo-schematic}\emph{d}-\ref{fig:HEalgo-schematic}\emph{f} show such cells with red circles at their centers. {Note that the saturation threshold $s_{w,T}$ needs to kept as close to unity as possible as it may otherwise alter the velocity of front propagation, but keeping it very close to unity can lead to very fast-moving interface as the saturation jump in Equation \eqref{eqn:vel_int} becomes very small, thus restricting the time step.} The degrees of freedom of faces corresponding to these saturated cells can be found using the discrete divergence operator. In case any subdomain(s) saturates completely the flux in the saturated region(s), $\textbf{q}_s$, is then evaluated by solving the elliptic problem \eqref{eqn:laplace-final} subject to Dirichlet boundary conditions \eqref{eq:pressure_bnd_cond-final}. Here we set the heads in all unsaturated cells and eliminate them using the projection approach discussed in Section~\ref{sec:BC-main}. This approach leads to a reduced system of equations corresponding only to the saturated cells (red circles) and hence efficiently and automatically deals with multiple, disconnected saturated regions. 

For saturated cells on the domain boundary $\partial \Omega_s \cap \partial \Omega $, shown by green lines in Figures \ref{fig:HEalgo-schematic}\emph{d}-\ref{fig:HEalgo-schematic}\emph{f}, the boundary condition specified on the external boundary must be applied. No flow boundary conditions are naturally implemented in the discrete gradient operator and therefore require no modification. For outflow boundary conditions the heads in the corresponding cells must be set to $h=- \int_\textbf{0}^\textbf{x}\hat{\textbf{g}} \cdot \d \textbf{x}$. 

Once the head is evaluated, the flux at the faces inside the saturated region (thin red lines) is evaluated using Darcy's law \eqref{eq:final-flux}, i.e., $\textbf{q}_s=-K \nabla h$. 
The fluxes on the cell faces corresponding to the saturated-unsaturated boundary, $\partial \Omega_s$ shown as a thick red line in Figure \ref{fig:HEalgo-schematic}, are upwinded according to the interface velocity \eqref{eqn:vel_int} as


\begin{align}\label{eq:shock-speed-criteria}
    \Flux_{w,~\textbf{x} \in \partial \Omega_s} = \begin{cases} \Flux_s, \quad \textbf{v}_{\textbf{x} \in \partial \Omega_s}\cdot\hat{\mathbf{n}} \geq 0, \\ \Flux_u, \quad \textbf{v}_{\textbf{x}  \in \partial \Omega_s} \cdot\hat{\mathbf{n}} < 0,\end{cases}
\end{align}
where $\textbf{v}_{\textbf{x} \in \partial \Omega_s}\cdot\hat{\mathbf{n}} > 0$ corresponds to local growth of the saturated region and $\textbf{v}_{\textbf{x} \in \partial \Omega_s}\cdot\hat{\mathbf{n}} < 0$ corresponds to a local contraction of the saturated region.
Domain boundaries which are saturated ($\partial \Omega \cap \partial \Omega_s$) as shown by a green line in Figure~\ref{fig:HEalgo-schematic}, only utilize the saturated flux $\textbf{q}_s$ unless a boundary condition is specified. Once the fluxes $\textbf{q}_s$ and $\textbf{q}_u$ are evaluated for saturated and unsaturated regions the total flux vector, $\textbf{q}_w^i$, can be assembled and used to update the saturation (or soil moisture content) explicitly from Equation \eqref{eq:discrete-mass-balance}. This coupled hyperbolic-elliptic PDE solution technique is summarized in Algorithm \ref{algo:hyperbolic-elliptic-PDE-solver}.

\begin{algorithm}
\caption{Hyperbolic-elliptic PDE solution algorithm}\label{algo:hyperbolic-elliptic-PDE-solver}
\begin{algorithmic}[1]
  \While{$t < T$} \Comment{Time loop}
\State Calculate \emph{gravity dominated flux} $\Flux_u$ ($=\Flux_w^i$) \eqref{eq:final-flux} at all cell faces in $\Omega$
\State Flag all the \emph{saturated} cells $\Omega_s$ where $s_w>s_{w,T}$ (threshold saturation) \Comment{$s_{w,T}\sim 1-s_{gr}$}
\If{$\Omega_s \neq \{  \}$} \Comment{Saturated region flux evaluation loop}
    \State Set Dirichlet boundary condition \eqref{eq:pressure_bnd_cond-final} on head, $h = - \int_\textbf{0}^\textbf{x}\hat{\textbf{g}} \cdot \d \textbf{x}$, in all unsaturated cells
    \State Set head, $h = - \int_\textbf{0}^\textbf{x}\hat{\textbf{g}} \cdot \d \textbf{x}$, in saturated cells on the domain boundary ($\partial \Omega_s \cap \partial \Omega$) to enable outflow
    \State Solve Laplace-type equation \eqref{eq:discrete-equation} {implicitly} for head $h$ 
    \State Calculate saturated region's face fluxes at all cell faces in $\Omega_s$ using Darcy's law \eqref{eq:final-flux}, i.e., $\Flux_s=-K\nabla h$
    \State Substitute the saturated cells' face fluxes in $\Flux_w^i$ considering the front motion criteria (\ref{eqn:vel_int} \& \ref{eq:shock-speed-criteria}) on $\partial \Omega_s$ \State \Comment{Domain boundaries which are saturated ($\partial \Omega \cap \partial \Omega_s$) only utilize saturated flux unless a boundary condition is specified}
\EndIf
\State \textbf{end if}
\State Calculate the time step $\Delta t^i$ from equation \eqref{eq:timestep}
\State Update the saturation $\textbf{s}_w^{i+1}$ \emph{explicitly} from Equation \eqref{eq:discrete-mass-balance}
\State Update time counter $t^{i+1}=t^i+\Delta t^i$
  \EndWhile
\State \textbf{end while}
\end{algorithmic}
\end{algorithm}

\section{Numerical tests} \label{sec:test-problems}
In this section we provide one and two dimensional, steady and transient test problems. We further validate our numerical solver against corresponding analytical results. For the examples considered in this section the absolute permeability, $\Ki(\phi)$, and relative permeability of water, $k_{rw}(s_w)$, follow power laws \citep{kozeny1927uber,carman1937fluid,brooks1964hydrau} given by
\begin{align} \label{eq:constitutive-fxn}
    \Ki(\phi) = \Ki_0 \phi^\m \quad \text{ and } \quad  k_{rw}(s_w)&=k_{rw}^0s_e^\n(s_w), \quad \text{ so that } \quad K(\phi,s_w) = \frac{\Ki_0 k_{rw}^0 \rho g}{\mu}\phi^\m s_e^\n(s_w) ,\\
    \text{with effective saturation }  s_e(s_w) &= \frac{s_w-s_{wr}}{1-s_{gr}-s_{wr}} \nonumber
\end{align}
where the coefficients $\Ki_0$ and $k_{rw}^0$, $\m$, $\n$ are model constants. Here the residual water and gas saturations are given by $s_{wr}$ and $s_{gr}$ respectively. The normalized relative permeability is $k_r(s_w,\n)=k_{rw}/k_{rw}^0=s_e^\n(s_w)$. The residual saturations are set to zero ($s_{wr}=s_{gr}=0$) in the numerical examples to match with the analytical results. Moreover, the power law exponents are set to $\m=3$ and $\n=2$ unless stated otherwise. The threshold for activation of the saturated region is $s_{w,T}\sim0.999$. The following dimensionless variables are introduced for depth, volumetric flux and time for sake of generality,

\begin{align}\label{eq:dimless-vars}
    z' = \frac{z}{z_0}, \quad \textbf{q}' = \frac{\textbf{q}}{K} \quad \text{and} \quad t' = \frac{t K}{z_0},
\end{align}
where $z_0$ is the characteristic depth (m) and $K$ is the saturated hydraulic conductivity (m/s) given by $K=K(\phi_0,1-s_{gr})$ with $\phi_0$ being the characteristic porosity of the problem. Typically, $\phi_0$ is the porosity at the surface ($z=0$) unless otherwise stated. For all the test cases presented in this section, the gravity vector, $\textbf{g}=(0,\,0,\,1)^T$, is aligned with the depth coordinate, $z$ and therefore the head in the unsaturated cells is $h=-z$.

\subsection{One-dimensional test cases}

\subsubsection{Drainage from a saturated soil} 
The first case is simple drainage of a fully-saturated soil with uniform porosity $\phi_0$. It corresponds to continued infiltration after the end of a heavy rainfall that has saturated the soil. It leads to formation of a spreading front called rarefaction wave \citep{Sisson1980}. This simple problem can help verify that the model captures the non-linearity of the unsaturated flow and the transition of the domain from saturated to unsaturated. The solution to this problem can be derived from the theory of hyperbolic equations \citep{lighthill1955kinematic,LeVeque1992,singh1997kinematic} and for the constitutive functions \eqref{eq:constitutive-fxn} we have as

\begin{figure}
    \centering
    \includegraphics[width=0.9\linewidth,trim = 0 0 0 0,clip]{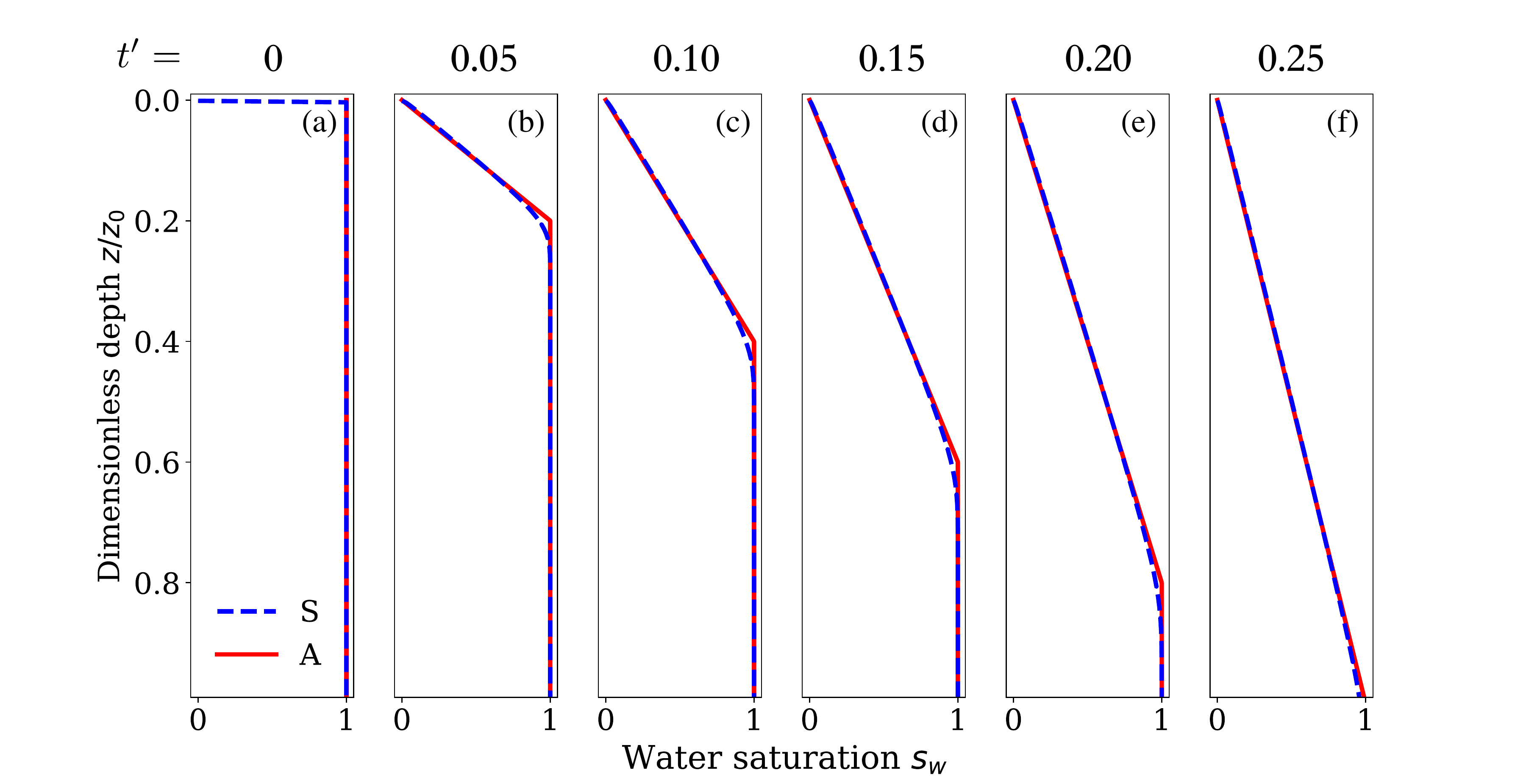}
    \caption{Drainage of an initially saturated, 50\% porous reservoir at different dimensionless times. The blue lines show the simulation results and the red lines show the analytic results. The numerical grid is 400 uniform cells on a domain with unit depth.}
    \label{fig:gravity-drainage}
\end{figure}

\begin{align}\label{eq:1D-drainage-dimless-sol}
    s_w(z',t') = \begin{cases} s_{wr}, \quad z' = 0\\  s_{wr} + \left(\frac{z' \phi_0}{\n t' } (1-s_{gr}-s_{wr})^\n\right)^{\frac{1}{\n-1}}, \quad  0 < \frac{z'}{t'} < \lambda' \\ (1-s_{gr}), \quad  \frac{z'}{t'} \geq \lambda' \end{cases}
\end{align}
where $\lambda'=\frac{\lambda}{K}= \frac{\n}{\phi_0  (1-s_{gr}-s_{wr})}$ is the dimensionless characteristic speed for the saturated domain moving downwards. Please note that for the constant porosity case, the dimensional flux is $q=K k_r(s_w,\n)$ and therefore the dimensionless solution \eqref{eq:1D-drainage-dimless-sol} is independent of $\m$.

Figure \ref{fig:gravity-drainage} shows drainage of an initially saturated porous reservoir which leads to the formation of a rarefaction wave traveling in the direction of gravity. Here the saturation linearly varies inside the rarefaction wave because the relative permeability depends quadratically on the saturation, i.e., $\n=2$. For this problem, the surface has a no-flow boundary condition whereas the base of the domain has an outflow boundary condition. The domain of unit depth is divided uniformly into 400 cells. The numerical solutions obtained from our proposed numerical method agree very well with the analytical results.

\subsubsection{Infiltration in a two-layered soil}\label{sec:two-layer-test}
\begin{figure}
    \centering
    \includegraphics[width=0.9\linewidth,trim = 0 0 0 0,clip]{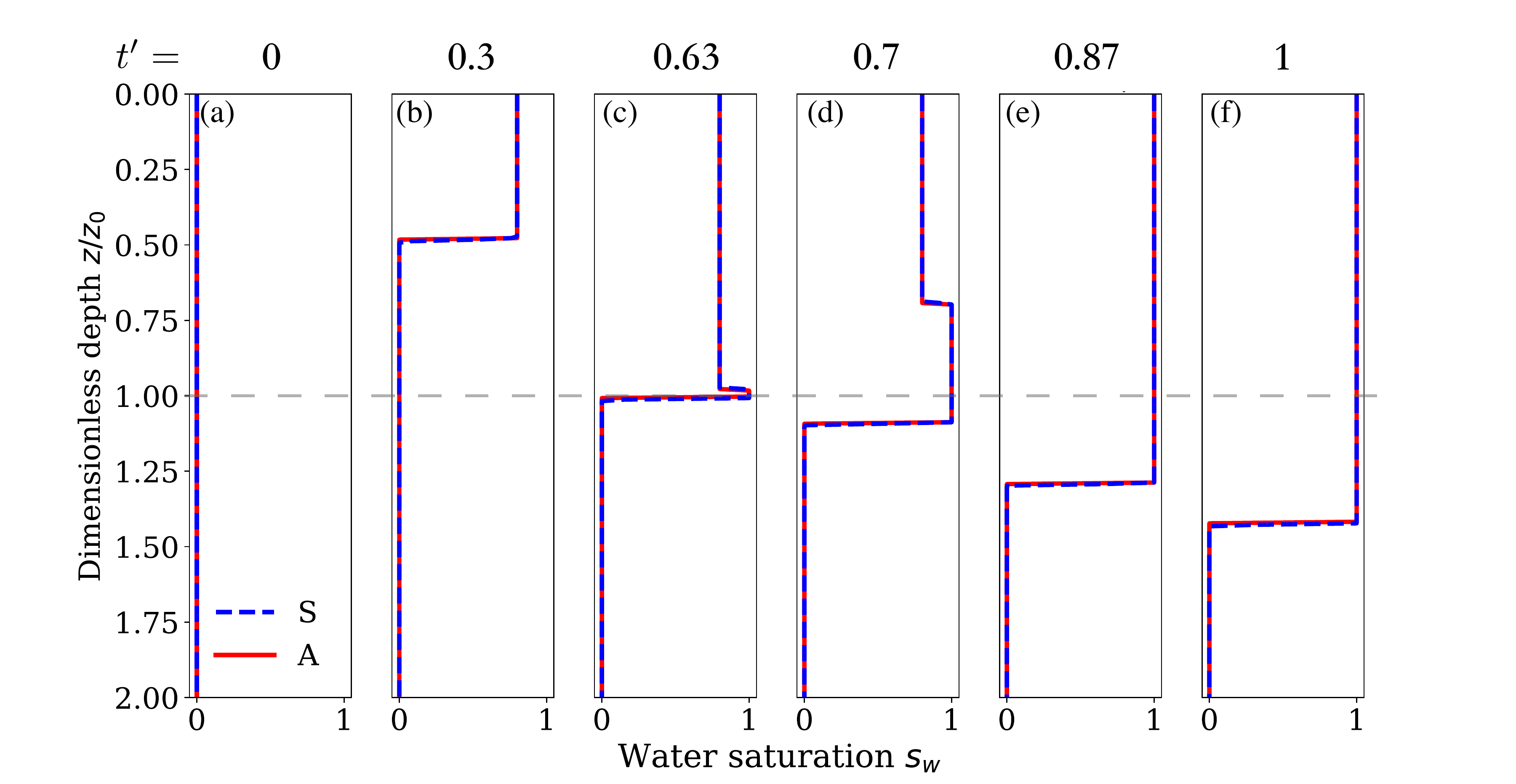}
    \caption{Infiltration process in an initially dry two-layered soil for $R/K_u=0.64$, $\phi_u=0.5$ and $\phi_l=0.2$ with a jump at $z'=1$ shown at different dimensionless times $t'=$ (a) 0, (b) 0.3, (c) 0.63, (d) 0.7, (e) 0.87 and (f) 1. The blue lines show the numerical solution (denoted as S) and the red lines show the analytical solution (denoted as A) given in \protect\cite{shadab2022analysis}. The dashed grey line at $z'=1$ ($z=z_0$) refers to the location of the jump.}
    \label{fig:two-layer-infiltration}
\end{figure}

The next 1D problem concerns infiltration in a dry, two-layered soil with a jump in porosity and conductivity at $z'=1$ as shown in Figure \ref{fig:two-layer-infiltration}\emph{a}. The porosities and conductivities of the upper and lower layers are $\phi_u,K_u$ and $\phi_l,K_l$, respectively. Here we consider the case of a more porous and conductive upper layer, so that $\phi_u>\phi_l$ and $\Ki_u>\Ki_l$. The soil is initially dry ($s_w(\textbf{x})=0$).

For transitional rainfall this leads to the spontaneous formation of a saturated region at the discontinuity and the propagation of two self-sharpening wetting fronts known as shock waves \citep{shadab2022analysis}. During transitional rainfall the rate of precipitation, $R$, is less than the conductivity of the upper layer but exceeds the conductivity of the lower layer ($K_l<R<K_u$). Initially, an unsaturated wetting front propagates downwards with constant velocity whose time-dependent location can be found using Rankine-Hugoniot jump condition (see Figure \ref{fig:two-layer-infiltration}\emph{b}) \citep{LeVeque1992}. Since the less porous layer is unable to accommodate the water flux as $K_l<R$, a fully saturated region forms at the location of the jump which grows outward, as shown in Figures~\ref{fig:two-layer-infiltration}\emph{c}-\ref{fig:two-layer-infiltration}\emph{d}. The initial wetting front thus bifurcates into two fronts moving upwards and downwards, bounding the saturated region. Once the upward moving shock reaches the surface, ponding occurs (see Figures~\ref{fig:two-layer-infiltration}\emph{c}-\ref{fig:two-layer-infiltration}\emph{d}). The locations of all the fronts can be found analytically using extended kinematic-wave approximation, as given in \cite{shadab2022analysis}.

For validation of the numerical method, we consider a dimensionless rainfall rate of $R/K_u=0.64$, in a dimensionless domain, $z'={z}/{z_0}\in[0,2]$, uniformly divided into 400 cells, with a porosity jump at $z'=1$. The porosity in upper region ($z'<1$) is $\phi_u=0.5$ whereas porosity in the lower region ($z' \geq 1$) is $\phi_u=0.2$. The numerical results agree well with the analytical results from \cite{shadab2022analysis}. The base of the domain is an outflow boundary, whereas the top is set to a constant saturation $s_w=0.8$ corresponding to $R/K_u=0.64$ as long as the flow is unsaturated. After the water ponds, the top becomes part of the saturated region, and a constant head equal to the gravitational head ($h=-z$) is applied.

{In Appendix \ref{sec:hydrus1Dcomparison}, we further compare the proposed numerical model against the solutions of the full Richards equation with capillary effects provided by the commonly used software Hydrus \citep{vsimunek2012hydrus}. At small scales where capillary gradients are fully resolved, both our model and Hydrus match analytical solutions for the wetting front  (see Appendix \ref{sec:smaller-scale} and Figure \ref{fig:12}). At large scales with low resolution, the full Richards solver in Hydrus fails to converge, either due to the sharp variations at the wetting fronts or unresolved capillary pressure variations (Figure \ref{fig:infiltration-figures}\textit{a}), whereas our solver robustly provides results that match analytic solutions (Figure \ref{fig:infiltration-figures}\textit{c}). 
If grid resolution is increased for this case, Hydrus converges but the solution does not conserve mass and the results are oscillatory near the wetting front, due numerical instabilities from the unresolved capillary pressure term (Figure \ref{fig:infiltration-figures}\textit{b}). In contrast, our solver continues to provide a robust monotone solution that resolves the wetting front more accurately (Figure \ref{fig:infiltration-figures}\textit{d}) {and conserves mass discretely. This is demonstrated by the mass balance ratio criteria, defined as the ratio of the total mass in the domain to the flux of water at the boundaries \citep{Celia1990}. A unit value of the mass balance ratio indicates conservation of mass, which is demonstrated by the present numerical technique for both medium and coarse grids (see Figure~\ref{fig:Massbalance}), where its variation is on the order of machine precision, i.e., $10^{-12}$.}

Lastly, when Hydrus converges it takes far more iterative steps (Figure \ref{fig:cum-iter}\textit{b}) compared to the present solver (Figure \ref{fig:cum-iter}\textit{d}). In addition, the iterations from our model are cheaper because we only solve a linear system for the cells in the saturated region while Hydrus solves a linear system for the whole domain. This reduction in computational cost increases in the multi-dimensional problems considered below.


\subsection{Two-dimensional steady test cases}
For the two-dimensional problems, the gravity is pointing vertically downwards, i.e., $\hat{\textbf{g}} \equiv (0,0, 1)^T$. The horizontal direction is scaled with the same characteristic length as the depth, so that $x' =x/z_0$, is added to the previously defined dimensionless variables, given in Equation \eqref{eq:dimless-vars}.

\subsubsection{Perched aquifer over a horizontal barrier}
This test concerns an idealized two-dimensional perched aquifer that develops when a constant discharge per unit depth, $Q_i$, infiltrates into a soil that includes a horizontal flow barrier of characteristic length, $L$ (Figure~\ref{fig:horizontal_baffles}\emph{a}). As the infiltrating fluid encounters this barrier it initially spreads laterally as a gravity current \citep{Huppert1995}. Once the fluid pours over the edges of the barrier the gravity current evolves to a steady shape. If the steady current is of low aspect ratio, with $L$ much less than its vertical extent, approximate analytic solutions can be found using the Dupuit approximation \citep{Hesse2010}. 

The gravity current divides the incident discharge, $Q_i$, into two discharges over the left and right edges, $Q_a$ and $Q_b$, located at distances $L_a$ and $L_b$ respectively from where the water impinges on the barrier (see Figure~\ref{fig:horizontal_baffles}\emph{a}). \citet{Hesse2010} use the Dupuit approximation to estimate the fraction ($f_b$) of the incident discharge $Q_i$ which spills over the right end of the barrier as
\begin{align}
    f_b = \frac{Q_b}{Q_a + Q_b} = \frac{L_a}{L_a + L_b}.
\end{align}
This theoretical result suggests that the discharge, $Q_b$, increases linearly with the distance of the source from the opposite edge, $L_a$. Comparison with experimental discharge measurements confirms this and shows good quantitative agreement with the theory (Figure~\ref{fig:horizontal_baffles}\emph{i}). 

For the numerical experiments, a 40\% porous domain $\Omega \equiv[0,7] \times[0,4]$ is divided into $140 \times 80$ cells. The impermeable barrier is placed in the region $[0.5,6.5] \times [3,3.3]$, and a no-flow boundary condition is imposed along its edge. At the top of the domain the incident discharge infiltrates over a thin region with a dimensionless width of 0.2 (Figure~\ref{fig:horizontal_baffles}\emph{e}), which results in a surface saturation of $s_w=0.975$. Boundary conditions are only required at the base of the domain and set to outflow. Numerical solutions are evolved to steady state for different locations of the incident discharge (Figures~\ref{fig:horizontal_baffles}\emph{e}-\ref{fig:horizontal_baffles}\emph{h}). Visual comparison with experimental images from \cite{Hesse2010} (Figures~\ref{fig:horizontal_baffles}\emph{a}-\ref{fig:horizontal_baffles}\emph{d}), show qualitative agreement of the current shapes. Further, an excellent quantitative comparison between the theoretical, experimental and simulation results can be observed for the dependence of discharge partitioning, $Q_a/(Q_a+Q_b)$, on the source location, $L_a/(L_a+L_b)$, as shown in Figure~\ref{fig:horizontal_baffles}\emph{i}.

\begin{figure}
    \centering
    \includegraphics[width=\linewidth,trim = 0 0cm 0 0,clip]{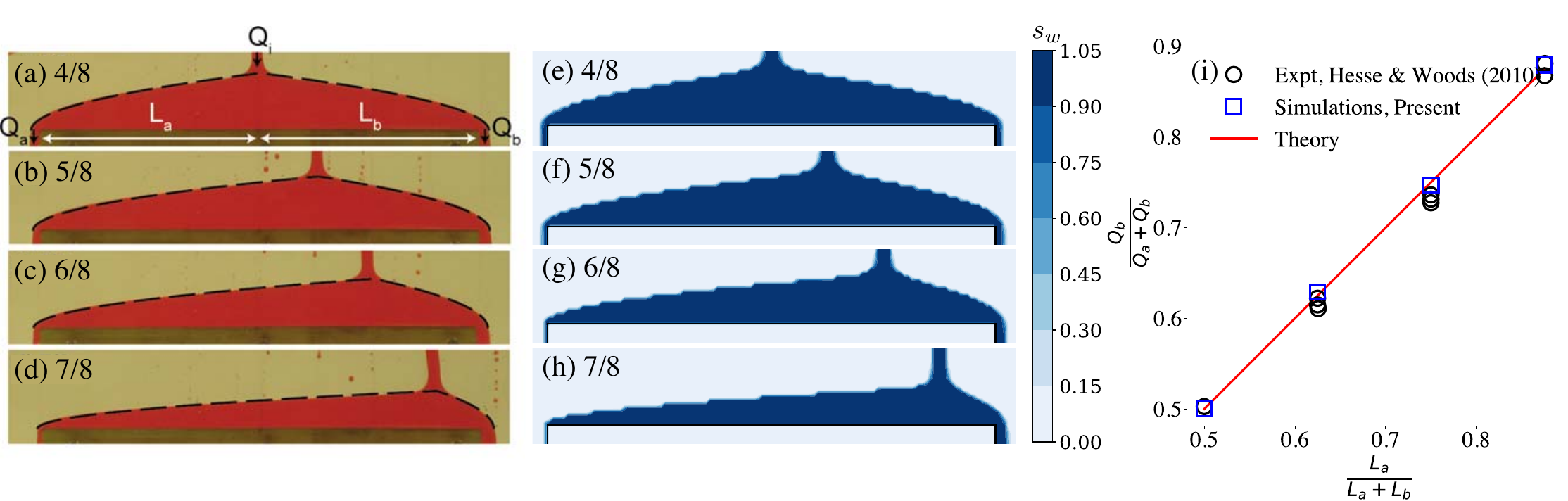}
    \caption{Flux partitioning in a steady gravity current passing over a barrier. (a-d): Experiments conducted by \protect\cite{Hesse2010} with glycerol in a Hele-Shaw cell of width 3 mm and a barrier of width 30 cm. (e-f): Simulations performed using the algorithm proposed in the present work. (i) Dependence of the flux partitioning, $Q_b/(Q_a+Q_b)$, on the source location, $L_a/(L_a+L_b)$. The panel labels give the dimensionless source locations, $L_a/(L_a+L_b)$. Note that simulation results in panels (e-h) do not show the entire computational domain to facilitate visual comparison with experiments.}
    \label{fig:horizontal_baffles}
\end{figure}

\subsubsection{Unconfined aquifer with a vertical seepage face}
Free surface flow in homogeneous porous reservoirs with development of a seepage face is a classic problem in subsurface hydrology \citep{Bear_1972,polubarinova2015theory,shadab2023investigating}.
Here we consider unconfined steady groundwater flow in a homogeneous 2D vertical cross-section of horizontal extent $L$ with an impermeable base. Fluid is injected at the left boundary and an outflow boundary with atmospheric pressure is applied on the right. In the resulting steady flow the groundwater table slopes to the right and its non-zero elevation at the right boundary is referred to as the height of the seepage face, $H_0$ (Figure~\ref{fig:seepage-face}\emph{a}). 

For the numerical simulations, the dimensionless domain $\Omega \equiv [0,1]\times[0,1]$ is divided into a uniform mesh of $75 \times 75$ cells. The medium has a porosity of 0.5 and is initially dry, i.e., $s_w(\textbf{x})=0$. The boundary condition on the left is a fixed discharge per unit width (m$^\textrm{2}$/s), $Q$, which is scaled as $Q/(K L)$. The boundary condition at the seepage face is dynamic until the steady state is achieved. In the unsaturated region above the groundwater table the boundary condition is not necessary and in the saturated region below a constant head equal to atmospheric pressure is applied, $h=-z$. No flow boundary conditions are applied at the top and bottom.

The approximate analytical solution to this problem is provided by \cite{polubarinova2015theory} but has to be evaluated numerically due to its complexity \citep{shadab4583340gui}. The numerical results for height of the groundwater table are very close to this analytical solution for $Q/(KL)=0.087$ and $0.211$ as shown in Figures~\ref{fig:seepage-face}\emph{a} and \ref{fig:seepage-face}\emph{b} respectively. Moreover, the variation in the dimensionless height of seepage face, $H_0/L$, with the dimensionless discharge per unit width, $Q/KL$, shows an excellent quantitative agreement in between the analytical and the numerical results (Figure~\ref{fig:seepage-face}\emph{c}).

\begin{figure}
    \centering
    \includegraphics[width=\linewidth,trim = 0 0cm 0 0,clip]{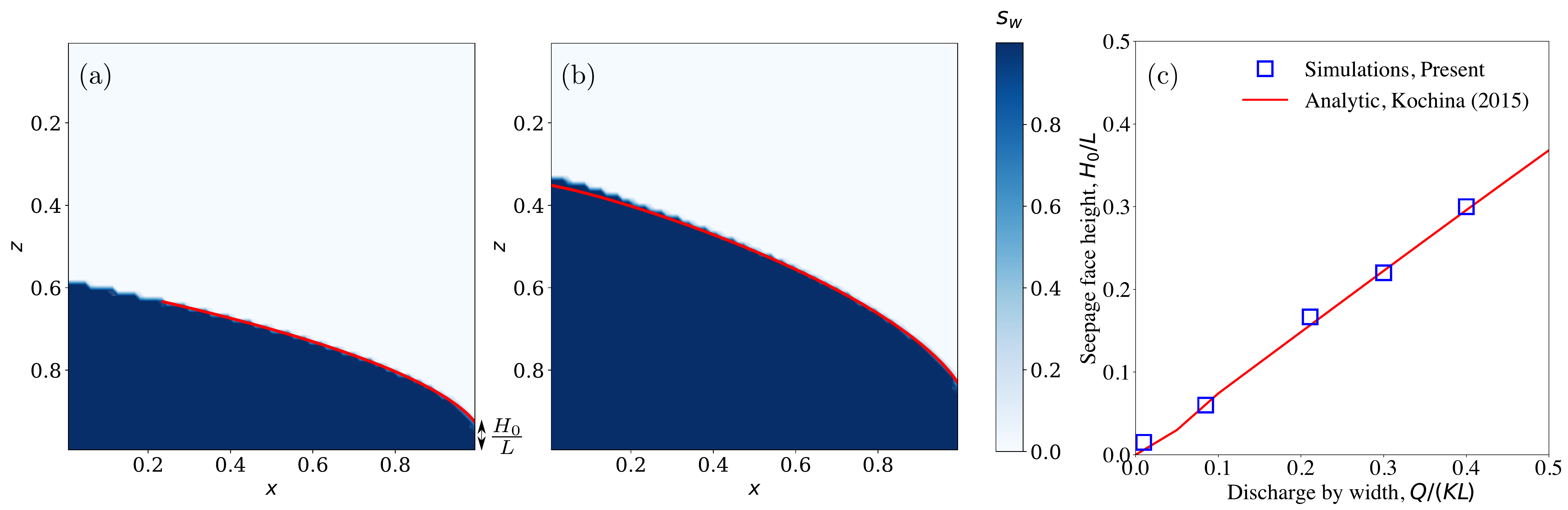}
    \caption{Steady unconfined aquifer with a vertical seepage face for dimensionless discharge per unit widths, $Q/(KL)=$ (a) 0.087 and (b) 0.211. (c) Variation of dimensionless seepage face heights, $H_0/L$, with the dimensionless discharge per unit width. The red lines are \protect\cite{polubarinova2015theory} analytic solutions evaluated from the software given in \protect\cite{shadab4583340gui}.}
    \label{fig:seepage-face}
\end{figure}

\subsection{Two-dimensional transient test cases}
No exact solutions to variably saturated, transient 2D flows are available, but for geometries with high aspect ratios, the Dupuit-Boussinesq theory provides a good approximation of the dynamics \citep{Troch2013}. Solutions can typically be found for simple scenarios where the shape of the water table is invariant under a stretching transformation \citep{Barenblatt1996}. These self-similar solutions, while approximate, typically compare very well with suitable experiments and predict the first-order dynamics \citep{Woods2014}.
Below we use two such self-similar solutions to compare with the transient numerical results for the evolution of an unconfined aquifer. 

\subsubsection{Fluid drainage from the edge of a horizontal aquifer}\label{sec:2Ddrainage-edge-transient}
The first transient test is on buoyancy-driven drainage of a horizontal aquifer \citep{Rupp2005,Zheng2013}. 
Consider a rectangular, homogeneous aquifer with porosity $\phi_0$ that is fully saturated initially. If all boundaries are no-flow except the right boundary ($x=L$) which is an outflow, the evolution of the groundwater table is self-similar at late time and in the Dupuit-Boussinesq approximation the problem reduces to a nonlinear diffusion equation \citep{Rupp2005}. This self-similar analysis neglects the effects of seepage face and vertical flow which is a good approximation if the aquifer is low aspect ratio where the horizontal extent of the aquifer is much larger than its vertical extent. Following \cite{Zheng2013}, we introduce the dimensionless variables $h' = h/L$ and  $\eta=x/L$ and assume the self-similar solution takes the form
\begin{align}
    h' =  \frac{\f(\eta) \phi_0}{t'}
\end{align}
where the dimensionless function $\textrm f(\eta)$ is defined by the boundary value problem
\begin{align}\label{eqn:drainage ode}
\frac{\d}{\d \eta}\left(\f \frac{\d \f }{\d \eta}\right) + \f = 0, \quad  \text{subject to} \quad  \frac{\d \f }{\d \eta}\bigg|_{0}=0 \quad \text{and} \quad \f(1)=0.
\end{align}
This differential equation can be solved numerically to obtain the self-similar shape of the current at a late time.

Moreover, the total volume of groundwater inside the domain at a late time is defined as $\mathcal{V}(t) = w\phi_0 \int_0^L h(x,t) \d x $ where $w$ is the depth in the third dimension. So the total dimensionless volume of fluid inside the domain at any time is $\mathcal{V}'(t')$ which is defined as
\begin{align}\label{eq:2Ddrainage-volume}
   \mathcal{V}'(t')=\frac{\mathcal{V}(t)}{w L^2 } = \frac{\phi_0^2 }{t'} \int_0^1 \f(\eta) \d \eta \propto\frac{1}{t'}.
\end{align}
The groundwater volume is therefore predicted to decline as $1/t'$ at late time. Below we refer to the solution of \eqref{eqn:drainage ode} as semi-analytical, because it can be solved to any desired accuracy with standard numerical integration techniques.

For the numerical simulation, a low aspect ratio domain $\Omega \equiv [0,3] \times [0,1]$ is considered to reduce the effect of vertical flow and the height of the seepage face. The domain is divided into $150 \times 100$ cells. The porosity of the medium is $0.5$. All boundaries are no flow except the right one which is an outflow. The medium is $90\%$ saturated initially however the initial condition does not affect the solution at late stages, because all initial conditions will asymptote to the self-similar solution at late time \citep{Barenblatt1996}.  The resulting gravity current drains from the edge of the reservoir as shown in Figures~\ref{fig:2D-drainage}\emph{a}-\ref{fig:2D-drainage}\emph{e}. Because the self-similar solution only approximates the late-stage behavior we show the current evolution for $t'/\phi_0\geq6$. At late times, numerical and semi-analytical solutions agree fairly well but the numerically computed groundwater table is slightly higher than the semi-analytical results. The difference in the results is likely due to the development of a seepage face at the edge (Figure~\ref{fig:seepage-face}), but may also be due to non-zero vertical flow neglected in the semi-analytic solution. This difference in the height of the groundwater table also affects the evolution of dimensionless volume shown in Figure~\ref{fig:2D-drainage}\emph{f}, where at late stages the simulated volume of fluid is more than the semi-analytical volume evaluated from \eqref{eq:2Ddrainage-volume}. But the numerical result asymptotes to the theoretical scaling, $\mathcal{V}' \sim t'^{-1}$. 

\begin{figure}
    \centering
    \includegraphics[width=\linewidth]{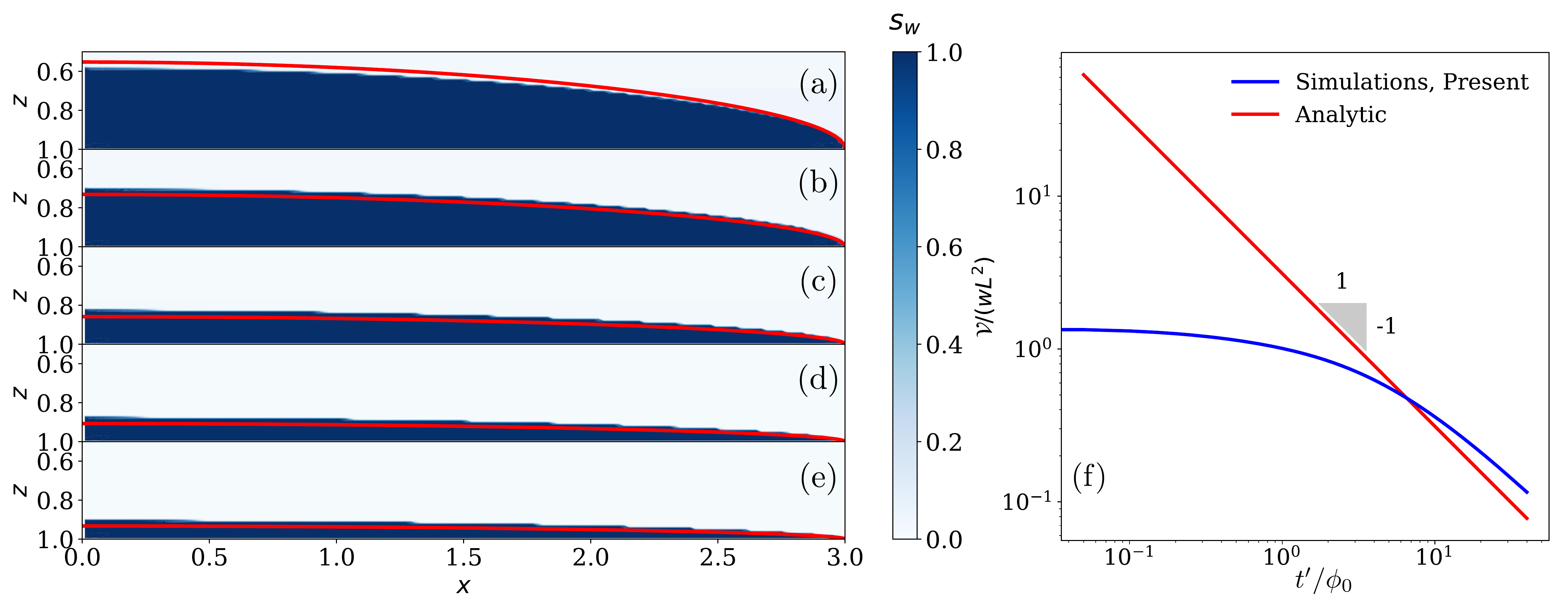}
    \caption{Drainage from the edge of horizontal aquifer at dimensionless times $t'/\phi_0=$ (a) $6$, (b) $12$, (c) $18$, (d) $24$ and (e) $30$. (f) The corresponding evolution of total dimensionless volume of fluid inside the domain, $\mathcal{V}/(wL^2)$. In all plots, the red lines show the semi-analytic solutions. The blue lines or contours show numerical solution.}
    \label{fig:2D-drainage}
\end{figure}

\subsubsection{Propagation of a gravity current into a porous layer}
Next we consider the propagation of a finite volume of groundwater into homogeneous horizontal porous medium that is initially dry. \citet{Hesse2007} have shown that at late time when the groundwater table is much lower than the upper boundary the flow asymptotes to classic self-similar solution found by \citet{Barenblatt1952}. \citet{Huppert1995} show that this solution, often referred to as a gravity current, provides a good approximation to experimental results. Consider the instantaneous release of finite volume of fluid, $\mathcal{V}$ (m$^\text{3}$), in a horizontal porous medium with constant porosity $\phi_0$. Following \citet{Huppert1995} the approximate analytic solution of the free-surface height, $h(x,t)$, is described using dimensionless variables introduced in section \ref{sec:2Ddrainage-edge-transient}. The dimensionless analytical solution is therefore given by
\begin{align}\label{eq:gravity-current}
    h'(x',t') = \left( \frac{\phi_0 \mathcal{V}'^2 }{t'} \right)^{\frac{1}{3}} f(\omega), \quad \text{with } f(\omega) = \frac{\omega_0^2 - \omega^2}{6}, \quad 0< \omega < \omega_0
\end{align}
where $\omega_0 = \left( \frac{9}{\phi_0} \right)^{1/3}$, $\omega = \frac{x'}{\left(\mathcal{V}' t'/\phi_0 \right)^{1/3}}$ and $\mathcal{V}'=\mathcal{V}/w z_0^2$, as introduced in Section \ref{sec:2Ddrainage-edge-transient}.
\begin{figure}
    \centering
    \includegraphics[width=\linewidth,trim = 0 0cm 0 0,clip]{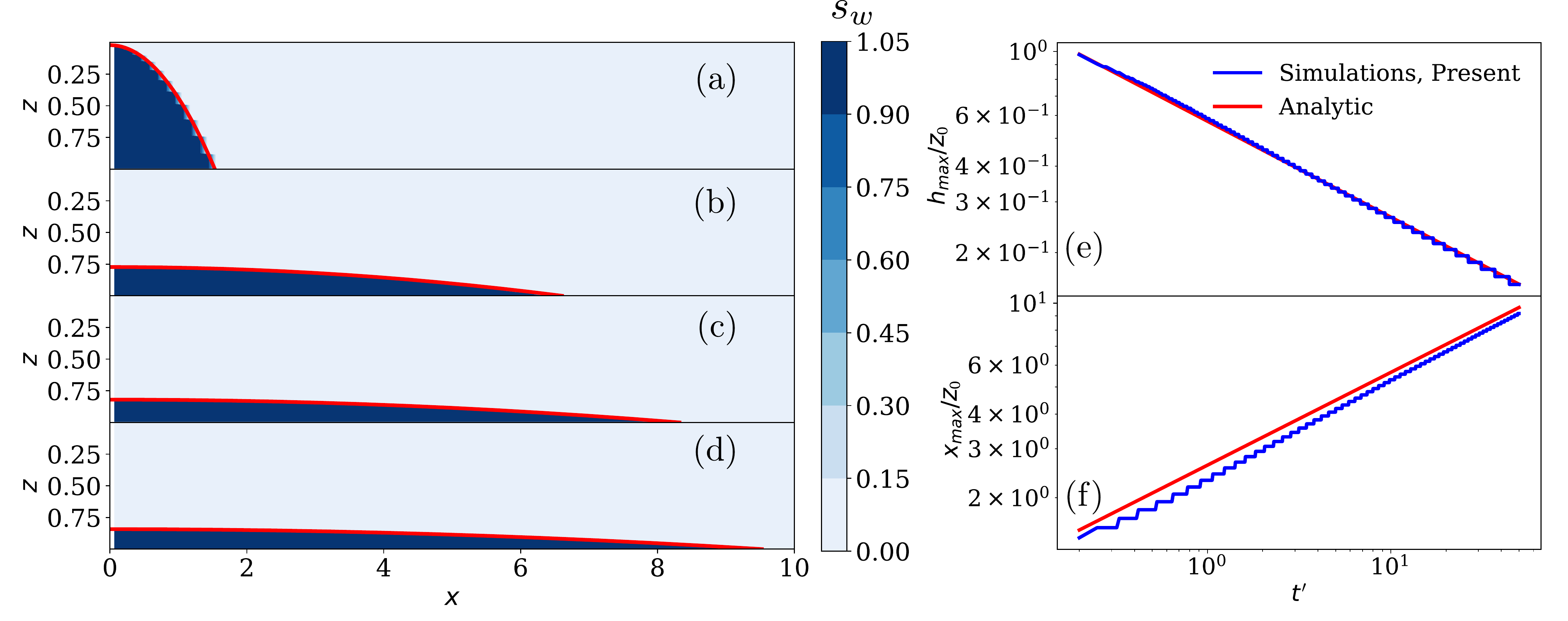}
    \caption{Gravity current propagation on a horizontal porous layer at dimensionless times $t'=$ (a) 0.2, (b) 16, (c) 32 and (d) 48. The spatial coordinates are dimensionless. The blue contour shows the solution from the proposed numerical method. The corresponding evolution of (e) dimensionless maximum height of the mound and its (f) dimensionless maximum spreading distance along x-axis is also provided. The solid blue line refers to the numerical solution. In all the plots, the red line shows the approximate analytic solution given by Equation \eqref{eq:gravity-current}. }
    \label{fig:horizontal_gravity-current}
\end{figure}

The 50\% porous domain $\Omega \equiv [0,25] \times [0,1]$ is divided into $200 \times 100$ cells. In this problem, the initial condition is set as the analytic solution at $t'=0.2$ (see Figure~\ref{fig:horizontal_gravity-current}\emph{a}). The boundary conditions are natural on left and bottom boundaries and are not required on top and right boundaries. The groundwater thus generated migrates towards right as shown in Figures~\ref{fig:horizontal_gravity-current}\emph{a}-\ref{fig:horizontal_gravity-current}\emph{d}. The numerical solutions are at par with the approximate analytical results given by Equation \eqref{eq:gravity-current} shown by red lines. It can be quantitatively observed in the time evolution of maximum height of the water table, $h'_{max}$, along $z$ axis and maximum spreading distance, $x'_{max}$, along the $x$ axis in Figures~\ref{fig:horizontal_gravity-current}\emph{e} and \ref{fig:horizontal_gravity-current}\emph{f} respectively. Although both numerical results follow the analytical results with the theoretical scaling ($h'_{max}\sim t'^{-1/3}$, $x'_{max}\sim t'^{1/3}$), there is a slight difference in the $x'_{max}$ results due to finite spatial resolution of the numerical simulations.

\subsubsection{Infiltration into a heterogeneous soil}\label{sec:heterogeneous}
The two-dimensional tests above have considered simple homogeneous media and have focused on the dynamics of the groundwater table. However, heterogeneity is a key characteristic of soils \citep{zhu2013characterizing} and the method proposed here can be applied to complex heterogeneous soils. Below we consider gravity-driven infiltration of rainwater into a heterogeneous soil and its interaction with multi-dimensional soil structure.

Here we consider a soil with porosity and permeability distributions given by correlated random fields (Appendix~\ref{sec:correlation}). We assume no spatial trend so the spatial correlation is only a function of the separation distance. Soils exhibit a {transverse anisotropy} (see \cite{zhu2013characterizing} for a review) where the horizontal scale of parameter fluctuation is often more than an order of magnitude larger than the vertical scale of fluctuation due to the nature of soil deposition \citep{yang2022algorithm}. 
Here we consider a stratified, heterogenous soil with two fluctuation lengths, $\theta_z=1$ and $\theta_x=10$ with the mean dimensionless permeability $\Ki_{mean}/\Ki_0=1$.  The resulting heterogeneous porosity profile is shown in Figure \ref{fig:heterogeneous-porosity-field}. 

This problem concerns rainwater infiltration in a layered soil in a domain $\Omega \equiv [0,5] \times [0,1]$ which is divided into $250 \times 100$ cells. The entire surface boundary condition ($z'=0$) is set to complete saturation ($s_w=1$), which corresponds to a heavy rainfall, in order to visualize the effects of multidimensional flow. The rest of the boundaries are set to be outflows. The resulting gravity driven infiltration is shown in Figure~\ref{fig:heterogeneous-field-solution} at different dimensionless times. Initially, an almost uniform front moves downwards (see Figures~\ref{fig:heterogeneous-field-solution}\emph{a}-\ref{fig:heterogeneous-field-solution}\emph{b}). But since the soil is more porous on the left half than the right half, the latter leads to formation of fully saturated region as shown in Figures~\ref{fig:heterogeneous-field-solution}\emph{c}-\ref{fig:heterogeneous-field-solution}\emph{d}. Whereas the front in the left half infiltrates faster. The water saturation on the left is lower in general as the medium is more porous as shown in Figure~\ref{fig:heterogeneous-field-solution}\emph{e}. Due to presence of lower porosity layer on the right half, the soil layers beneath remain inaccessible for very long periods of time. Inside the fully saturated regions, the dynamics switches from gravity driven flow to pressure driven flow. Once a saturated region forms, the flow can move in directions other than the gravity vector, i.e., laterally or upwards leading to ponding and runoff. In this problem multiple saturated (perched) regions form inside the domain but they are treated simultaneously and efficiently in the approach proposed here. This is a challenging problem due to soil heterogeneity as well as formation of multiple saturated regions.

\begin{figure}[tbp!]
    \centering
    \includegraphics[width=0.8\linewidth]{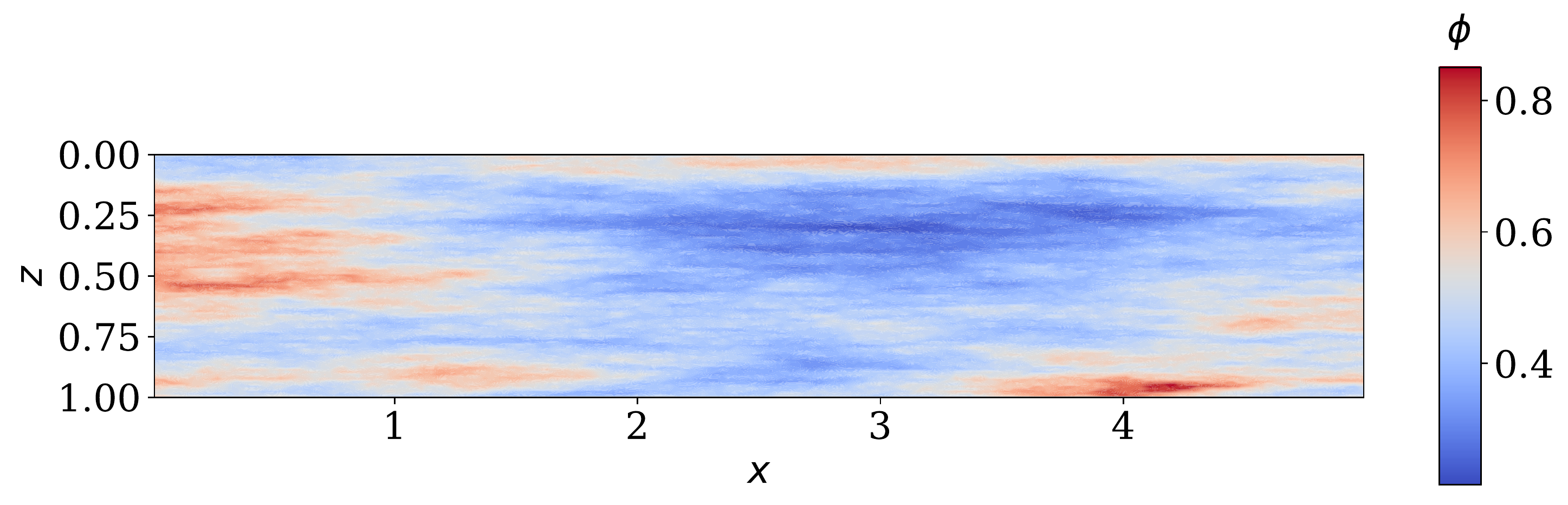}
    \caption{Porosity field of a stratified, heterogeneous soil with mean porosity  $\phi_{mean}=46.5\%$, maximum porosity $\phi_{max}=85\%$ and minimum porosity $\phi_{min}=21.6\%$ with fluctuation lengths $\theta_x=10$ and $\theta_z=1$.}
    \label{fig:heterogeneous-porosity-field}
\end{figure}
\begin{figure}[tbp!]
    \centering
    \includegraphics[width=0.8\linewidth]{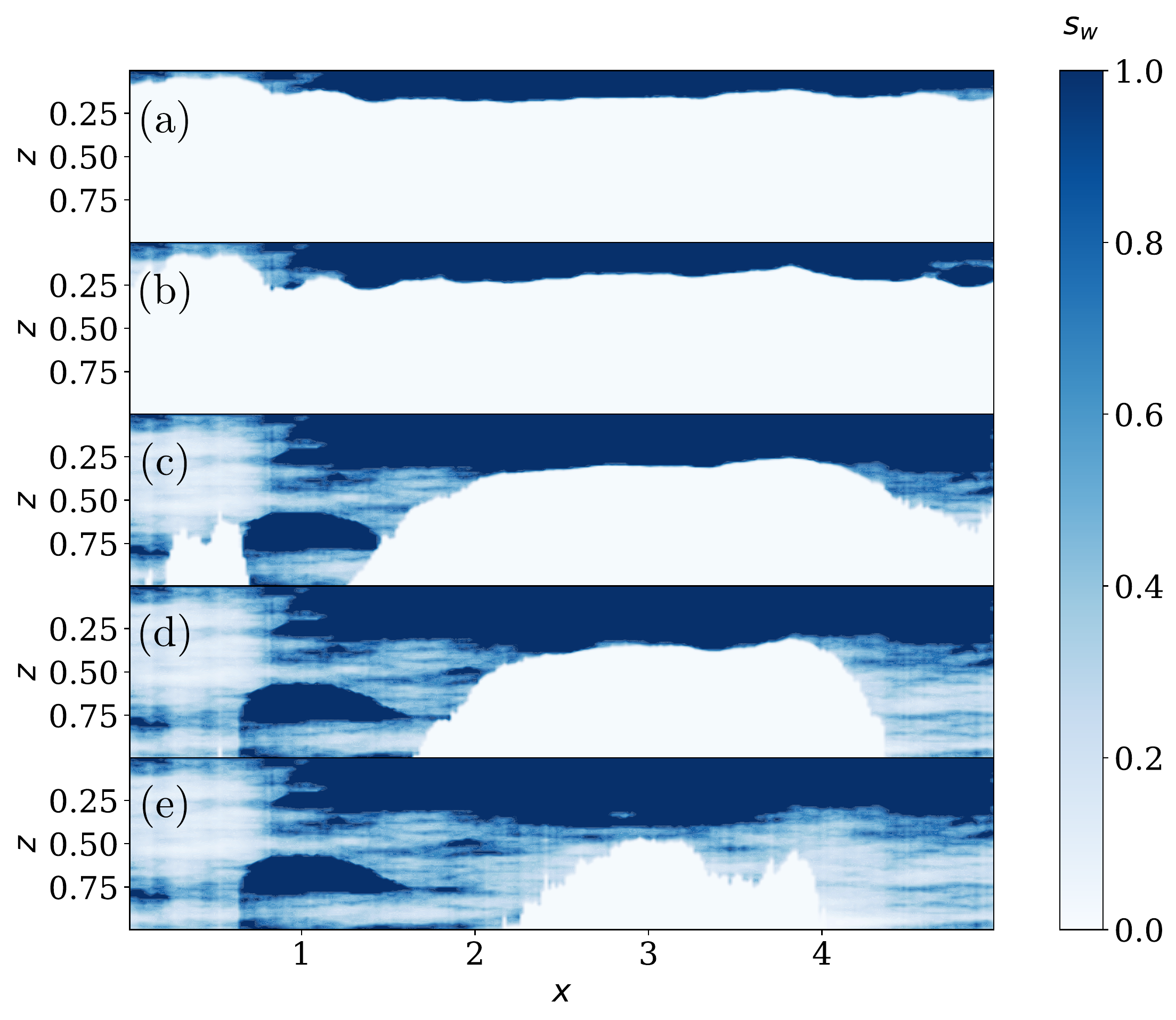}
    \caption{Time sequenced images of infiltration in a stratified, heterogeneous soil at dimensionless times $t'=$ (a) 5, (b) 10, (c) 50, (d) 100 and (e) 250.}
    \label{fig:heterogeneous-field-solution}
\end{figure}

\section{Conclusions}\label{sec5:conclusions}
{In this paper we first introduced a hyperbolic-elliptic partial differential equation based model for variably saturated groundwater flow in the limit of negligible capillary forces. The assumption of negligible capillarity holds either at large spatial scales and/or in low-textured soils such as sand.}
The technique switches dynamically from solving a hyperbolic problem for two-phase flow in unsaturated regions to solving an elliptic problem for single-phase flow in saturated regions. The developed multidimensional numerical model is based on a conservative finite difference scheme and tensor-product grid approach which can efficiently and robustly handle sharp saturation gradients as well as fully saturated regions. 
We also provide a suite of challenging benchmark problems in one and two dimensions along with their corresponding  (semi-) analytical results. These problems involve variably saturated flow which can help in verification and validation as well as performance comparison of numerical solvers. Our simulation results show excellent agreement with the analytical solutions for all the proposed problems. Finally, we consider a complicated test of rainwater infiltration into a stratified, heterogeneous soil. This test illustrates how the proposed method deals with the intricate infiltration dynamics which involves formation and evolution of multiple saturated regions also causing perching and lateral flow.

\appendix
\section{One dimensional operators}\label{sec:one-D-operators}
Here we build on the discrete operator approach initially outlined by \cite{Hesse2018a} and implemented in \cite{MDOTv1}. We extend the approach to include advection and three dimensions and describe the implementation of boundary conditions through the elimination of constraints in detail.

The gradient, divergence and mean operators are approximated with a second-order finite difference approximation. For a one-dimensional grid with $n$ ($m$ ; $l$) cells of uniform width $\Delta x$ ($\Delta z$ ; $\Delta y$) and therefore $(n+1)$ faces, the $n$ by $(n+1)$ discrete divergence operator, $\textbf{D}_{n}\in \mathbb{R}^{n \times (n+1)}$, and the $(n+1)$ by $n$  discrete gradient operator, $\textbf{G}_n \in \mathbb{R}^{(n+1)\times n} $, respectively take the forms

\begin{align}\label{eq:one-D-operators}
    \nabla \cdot \approx \textbf{D}_{n} = \frac{1}{\Delta x}\begin{bmatrix}
    -1 & 1 &  &  &  &  \\ 
     & -1 &1  &  &  & \\
     &  & \ddots & \ddots &  & \\
     &  &  & -1&1  &  \\
     &  &  & & -1 &1 
 \end{bmatrix}_{n~\times~(n+1)} \text{and} \quad 
    \nabla  \approx \textbf{G}_{n} = \frac{1}{\Delta x} \begin{bmatrix}
    0 &  &  &  &  &  \\ 
    -1 & 1 &  &  &  &  \\ 
     & -1 &1  &  &  & \\
     &  & \ddots & \ddots &  & \\
     &  &  & -1&1  &  \\
     &  &  & & -1 &1 \\
          &  &  & & & 0
 \end{bmatrix}_{(n+1)~\times~n}.
\end{align}
 The discrete gradient, $\textbf{G}_n$, assumes a natural boundary condition, i.e., no flow across the domain boundary. In a similar fashion, the $(n+1)$ by $n$ arithmetic mean operator, $\textbf{M}_{n}\in \mathbb{R}^{(n+1)\times n}$, can be constructed as 
\begin{align}\label{eq:one-D-mean}
    \textbf{M}_{n} = \frac{1}{2} \begin{bmatrix}
    m_b &  &  &  &  &  \\ 
    1 & 1 &  &  &  &  \\ 
     & 1 &1  &  &  & \\
     &  & \ddots & \ddots &  & \\
     &  &  & 1&1  &  \\
     &  &  & & 1 &1 \\
          &  &  & & & m_b
 \end{bmatrix}_{(n+1)~\times~n}.
\end{align}
Here $m_b=2$ for placing the boundary cell value at the corresponding boundary face without interpolating and $m_b=0$ for a zero value (no-flow boundary condition) as placeholders. The entries corresponding to the boundaries are typically replaced by the boundary conditions which will be discussed in Appendix \ref{sec:BC-full}.

\textit{Advection operator}. For solving the hyperbolic Richards equation, the advective flux computation requires value of saturation (or soil moisture content) at the faces to evaluate the hydraulic conductivity. For that purpose, we use Darcy flux-based upwinding to compute the face values of saturation \citep{godunov1959finite}. In matrix form, we construct an advection matrix operator, $\textbf{A}(\textbf{q}_w)\in \mathbb{R}^{(n+1)\times n}$, which takes in the $(n+1)$ face flux vector $\textbf{q}_w=[q_1,q_2,\hdots,q_{n-1},q_{n+1}]^T$ and provides the $(n+1)$ by $n$ advection operator \textbf{A}. The one-dimensional advection operator $\textbf{A}$ is constructed as

\begin{align}
    \textbf{A}_n = \textbf{Q}_{(n+1)}^+\textbf{A}_n^+ ~+~ \textbf{Q}_{(n+1)}^-\textbf{A}_n^- 
\end{align}
where in one dimension, 

\begin{align}\label{eq:one-D-advection-operators}
    \textbf{Q}_{(n+1)}^+ &=  \begin{bmatrix}
    q_1^+ &  &  &  &  &  \\ 
    &  q_2^+ &  &  &  &  \\ 
     & & q_3^+  &  &  & \\
     &  &  & \ddots &  & \\
     &  &  &  & q_{n}^+  &  \\
     &  &  & & & q_{n+1}^+ \\
 \end{bmatrix}_{(n+1)~\times~(n+1)}, \quad    & \textbf{Q}_{(n+1)}^- =  \begin{bmatrix}
    q_1^- &  &  &  &  &  \\ 
    &  q_2^- &  &  &  &  \\ 
     & & q_3^-  &  &  & \\
     &  &  & \ddots &  & \\
     &  &  &  & q_{n}^-  &  \\
     &  &  & & & q_{n+1}^- \\
 \end{bmatrix}_{(n+1)~\times~(n+1)},  \\
    \textbf{A}_n^+ &=   \begin{bmatrix}
    0 &  &  &  &  &  \\ 
    1 &  &  &  &  &  \\ 
     & 1 &  &  &  & \\
     &  & \ddots &  &  & \\
     &  &  & 1&  &  \\
     &  &  & & 1 & \\
          &  &  & & & 1
 \end{bmatrix}_{(n+1)~\times~n} \quad \quad \quad 
\text{and} \quad   &   \textbf{A}_n^- =   \begin{bmatrix}
    1 &  &  &  &  &  \\ 
     & 1 &  &  &  &  \\ 
     &  &1  &  &  & \\
     &  &  & \ddots &  & \\
     &  &  & &1  &  \\
     &  &  & &  &1 \\
          &  &  & & & 0
 \end{bmatrix}_{(n+1)~\times~n},
\end{align}
where $q_i^-$ refers to negative flux, $\min\{q_i,0\}$, and $q_i^+$ refers to positive flux, $\max\{q_i,0\}$, for $i \in \{1,2,\hdots,n, n+1\}$.

\section{Discrete operators in three dimensions} \label{sec:3Dextension}
The present framework can be easily extended to three dimensions with the same approach. The tensor product grid enables a straightforward extension from two to three dimensions with a few more lines of code. In three dimensions, we use a regular Cartesian mesh with $n$ cells of size $\Delta x$ in $x$ direction, $m$ cells of size $\Delta y$ in $y$ direction and $l$ cells of size $\Delta z$ in $z$ direction. The three-dimensional divergence, gradient and mean operators ($\textbf{D},\textbf{G}$ and $\textbf{M}$ respectively) are then composed of three block matrices as

\begin{align}
  \textbf{H} = \begin{bmatrix}\textbf{H}_x \\\textbf{H}_y\\\textbf{H}_z \end{bmatrix},\quad \textbf{H}\in \{\textbf{D}^T~,~\textbf{G}~,~\textbf{M}\}
\end{align}
with directional operators defined as
\begin{align}\label{eq:three-D-operators}
    \textbf{H}_x = \textbf{I}_l \otimes(\textbf{H}_n \otimes \textbf{I}_m), \quad &  \quad \textbf{H}_y =  \textbf{I}_l \otimes (\textbf{I}_n \otimes \textbf{H}_m) \quad \text{and} \quad \textbf{H}_z =  \textbf{H}_l \otimes (\textbf{I}_n \otimes \textbf{I}_m) ,\quad \textbf{H}\in \{\textbf{D}~,~\textbf{G}~,~\textbf{M}\}.
\end{align}
Similar to two dimensional operators given in Section \ref{sec:discrete-operator}, the one dimensional operators, $\{\textbf{D}_\alpha,\textbf{G}_\alpha,\textbf{M}_\alpha\}$ for $\alpha \in \{ n,m,l\}$, can be evaluated from Equations~(\ref{eq:one-D-operators}-\ref{eq:one-D-mean}). Moreover, the $N_f$ by $N$ {advection operator} $\textbf{A}$ can be expressed in three dimensions as 

\begin{align}\label{eq:three-D-advection-operators1}
    \textbf{A} = \textbf{Q}^+\textbf{A}^+ ~+~ \textbf{Q}^-\textbf{A}^- ,
\end{align}
where $N_f$ is the total number of faces which is the summation of $x$, $y$ and $z$ normal faces, i.e., $N_f=N_{fx}+N_{fy}+N_{fz}$ and $N$ is the total number of cells given by $N=nml$. Additionally, the three-dimensional block matrix of positive or negative fluxes, $\textbf{Q}^\pm$, and the advection operator, $\textbf{A}^\pm$, are given by

\begin{align}\label{eq:three-D-advection-operators-sub}
\textbf{Q}^\pm =  \begin{bmatrix}
    \textbf{Q}_x^\pm & &\\
    & \textbf{Q}_y^\pm& \\
       & & \textbf{Q}_z^\pm
 \end{bmatrix}_{N_f~\times~N_f} \quad  \text{and} \quad  & \textbf{A}^\pm = \begin{bmatrix} \textbf{A}_x^\pm\\\textbf{A}_y^\pm\\ \textbf{A}_z^\pm \end{bmatrix}_{N_f~\times~N} .
 \end{align}
 In three-dimensions, the number of $x$, $y$ and $z$ normal faces are $N_{fx}=(n+1)ml$, $N_{fy}=n(m+1)l$ and  $N_{fy}=nm(l+1)$ respectively. $\textbf{Q}_x^\pm$ ($\textbf{Q}_y^\pm$ ; $\textbf{Q}_z^\pm$) is the $N_{fx}$ by $N_{fx}$ ($N_{fy}$ by $N_{fy}$ ; $N_{fz}$ by $N_{fz}$) diagonal matrix with $x$ ($y$ ; $z$) positive/negative face Darcy fluxes in the diagonal. Moreover,  the $x$, $y$ and $z$ matrix components of $\textbf{A}^+$ and $\textbf{A}^-$ are expressed similar to the 3D discrete operators \eqref{eq:three-D-operators} as follows,

\begin{align}\label{eq:three-D-advection-operators2}
     \textbf{A}_x^\pm = \textbf{I}_l \otimes \left(\textbf{A}_n^\pm \otimes \textbf{I}_m \right), \quad \quad \textbf{A}_y^\pm = \textbf{I}_l \otimes \left(\textbf{I}_n \otimes \textbf{A}_m^\pm \right) & \quad \text{and} \quad \textbf{A}_z^\pm = \textbf{A}_l^\pm \otimes \left(\textbf{I}_n \otimes \textbf{I}_m \right)
\end{align}
where $\textbf{A}_n^\pm$, $\textbf{A}_m^\pm$ and  $\textbf{A}_l^\pm$ are again the one-dimensional operators as expressed in \eqref{eq:one-D-advection-operators}. For the definitions of one-dimensional operators, see Equations~(\ref{eq:one-D-operators}-\ref{eq:one-D-mean}). The sequence is chosen to respect the internal ordering of Matlab where the cells and faces are ordered in $y$ direction first, then $x$ and then $z$. Please note that the vertical dimension in this case $y$ instead of $z$ as proposed earlier in two-dimensional cases for sake of consistency with the literature.

\section{Boundary conditions}\label{sec:BC-full}
Boundary conditions are required so that the PDE problem becomes well-posed. In this section, we discuss the implementation of the boundary conditions taking the discrete system \eqref{eq:discrete-equation} as an example. However, its application is more general which includes the discrete saturation equation \eqref{eq:discrete-mass-balance}. {Natural boundary conditions} (zero gradient or no flow) are already implemented in the discrete gradient operator so there is no extra effort. Note that if gradient is not being used in the evaluation of the flux, the natural boundary condition is enforced in the divergence operator by calculating it from $\textbf{D}=-\textbf{G}^T$. 

\textit{Neumann boundary condition}.
Its implementation is fairly straightforward. For zero flux at the boundary, nothing needs to be done since it is in-built in the discrete gradient. Although, the non-zero fluxes at the boundary are converted to an additional source term corresponding to Neumann boundary cells

\begin{equation}
    f_n = q_b \frac{A}{V} \label{eq:43neu}
\end{equation}
where $f_n$ is the source term due to the Neumann BC, $q_b$ is the boundary flux, $A$ is the area of the corresponding boundary face and $V$ is the volume of the corresponding boundary cell. The extra source term for Neumann boundary takes the form of $N$ by $1$ vector, $\textbf{f}_n$, which is added to the governing discrete equation. For example, the governing discrete equation \eqref{eq:discrete-equation} becomes $\textbf{L}~\textbf{h} = \textbf{f}_s+\textbf{f}_n$.

\textit{Dirichlet boundary conditions}.
Since Dirichlet boundary conditions prescribe the unknowns (for example, head $h$) at the boundary cell centers, the number of unknowns thus has to be reduced in accordance with the constraints. The two cases are considered as follows: \\

\textit{{a. Homogeneous Dirichlet BC}}: In this case, the system of equations has to be reduced according to the constraints. The idea is to project the solution vector, $\textbf{h} \in \mathbb{R}^N$, onto a reduced subspace that lives in $\mathbb{R}^{(N-N_c)}$ using an orthogonal projection, eliminating the $N_c$ number of constraints \citep{trefethen1997numerical}. The subspace of the reduced unknowns is the null space of the $N_c$ by $N$ constraint matrix \textbf{B}, i.e., $\mathcal{N}(\textbf{B}) \in \mathbb{R}^{(N-N_c)} $, where $\textbf{B}$ is defined as $\textbf{B}~\textbf{h}=\textbf{0}$. Any orthogonal projector made from a set of orthonormal bases for $\mathcal{N}(\textbf{B})$ can project the unknowns from full space in $\mathbb{R}^N$ to the reduced space $\mathcal{N}(\textbf{B}) \in \mathbb{R}^{(N-N_c)}$. Therefore, an $N$ by $ (N-N_c)$ matrix \textbf{N} can be constructed trivially with standard orthonormal bases with the orthogonal projector being $\textbf{N}^T  \hspace{1mm}\textbf{N}$ or $\textbf{N} \hspace{1mm} \textbf{N}^T$ which lives in $\mathbb{R}^{(N-N_c) \times (N-N_c)}$ \citep{trefethen1997numerical}. From an identity matrix $\textbf{I}_N$, \textbf{B} and \textbf{N} can be easily constructed by eliminating the columns for the former and keeping the rows for the latter, corresponding to the degree of freedom of the constraints. Further, $\textbf{L}_r=\textbf{N}^T \textbf{L} \hspace{1mm} \textbf{N}$ is defined as the discrete Laplacian operator matrix of reduced dimensions $(N-N_c) \times (N - N_c)$, $\textbf{h}_r=\textbf{N}^T \hspace{1mm} \textbf{h}$ is defined as the $(N-N_c)$ by $1$ vector of unknowns in the reduced subspace that belongs to $\mathbb{R}^{(N-N_c)}$ and $\textbf{f}_{s,r}=\textbf{N}^T \hspace{1mm} \textbf{f}_s $ is the $(N-N_c)$ by $1$ vector of the source term in the reduced subspace that lives in $\mathbb{R}^{(N-N_c)}$. The final reduced matrix equation is $\textbf{L}_r \textbf{h}_r = \textbf{f}_{s,r}$ which can be solved directly for $\textbf{h}_r$.

\textit{{b. Heterogeneous Dirichlet BC}}: Dealing with non-zero Dirichlet boundary conditions is slightly more sophisticated. The (quasi-)linearity of the problem helps split the solution $\textbf{h}$ into homogeneous $\textbf{h}_0$ and particular $\textbf{h}_p$ solution vectors of size $N$ by $1$ for solving the boundary value problem \citep{greenberg2013foundations}. The $N_c$ by $N$ constraint matrix $\textbf{B}$ is the same as earlier which governs $\textbf{B}~ \textbf{h}_p = \textbf{c}$ where $\textbf{c}$ is the $N_c$ by $1$ vector of known values at the Dirichlet boundaries.

\begin{equation}
\left.\begin{aligned}
\textbf{B} \hspace{2mm} \textbf{h}_0 = \textbf{0} \\
\textbf{B} \hspace{2mm} \textbf{h}_p = \textbf{c}
\end{aligned}\right\rbrace \hspace{2mm} \textbf{B} \hspace{1mm} (\textbf{h}_0+\textbf{h}_p)=\textbf{c} \label{eq:43constraint}
\end{equation}

Since \textbf{B} is a $N_c$ by $N$ right invertible matrix consisting of orthonormal bases, an orthogonal projector $\textbf{B}\hspace{1mm}\textbf{B}^T\in \mathbb{R}^{N_c \times N_c}$ of dimension $N_c\times N_c$ can be constructed \citep{trefethen1997numerical}. We define the particular solution $\textbf{h}_{p,r}$ in reduced subspace, that lives in $\mathbb{R}^{N_c}$, to be $
    \textbf{h}_{p,r} = \textbf{B} ~ \textbf{h}_p = \textbf{c}$
    and $\textbf{h}_{p} = \textbf{B}^T ~ \textbf{h}_{p,r}$.
Therefore, we solve $\textbf{B}^T ~\textbf{B}  ~ \textbf{h}_{p,r} = \textbf{c}$ to get $\textbf{h}_p$ from the Dirichlet boundary constraint matrix, \textbf{B}, and the Dirichlet boundary condition values, \textbf{c}.

Next, to find the homogeneous solution $\textbf{h}_0$, plugging the decomposed solution (\ref{eq:43constraint}) in the discrete governing equation \eqref{eq:discrete-equation} gives the final system, $\textbf{L} ~ \textbf{h}_0 = \textbf{f}_s + \textbf{f}_D$, where $\textbf{f}_D$ is source term from heterogeneous Dirichlet BCs,  $\textbf{f}_D=-\textbf{L}~\textbf{h}_p$. Then similar to the homogeneous Dirichlet boundary condition case, the constraints are eliminated using orthogonal projection, again leading to a well-posed, reduced system of equations which can be solved for the homogeneous solution $\textbf{h}_0$. Finally, the solution can be evaluated from $\textbf{h}=\textbf{h}_0+\textbf{h}_p$. 

\section{Generation of correlated random fields}\label{sec:correlation}
In Section~\ref{sec:heterogeneous} we use a correlated random field for both the porosity and the permeability in the heterogeneous test case. Assuming the scale of formation of fluctuation is elliptical, the exponential correlation function, $\mathcal{\varrho}$, for transverse anisotropy is given in \cite{zhu2013characterizing} as 

\begin{align}\label{eq:correlation-fxn}
    \varrho = \exp \left(- 2 \sqrt{\frac{\Delta \mathcal{X}^2}{\theta_x^2} + \frac{\Delta \mathcal{Z}^2}{\theta_z^2}} \right)
\end{align}
where $\Delta \mathcal{X}$ and $\Delta \mathcal{Z}$ are respectively the horizontal and vertical separation distances between two observations in the space. The symbols $\theta_x$ and $\theta_z$ denote the principal scales of fluctuation in the $x$ and $z$ directions respectively.

The matrix decomposition method is utilized in this approach which requires the defining a discrete set of $N$ spatial points at which the random field is sampled. Subsequently, it helps construct an $N$ by $N$ covariance matrix, $\textit{\textbf{C}}$ which quantifies the correlation between all of the spatial points being sampled. The exponential correlation function $\varrho$ \eqref{eq:correlation-fxn} helps define the covariance matrix. For generating realizations of correlated random field, matrix decomposition method is often utilized \citep{zhu2013characterizing}. An expensive but exact method is Cholesky factorization \citep{trefethen1997numerical} which is performed on the covariance matrix to obtain upper ($\textit{\textbf{L}}^T$) and lower ($\textit{\textbf{L}}$) triangular matrices as

\begin{align*}
    \textit{\textbf{L}}~ \textit{\textbf{L}}^T = \textit{\textbf{C}}.
\end{align*}
Next an $N$ by $1$ vector $\textit{\textbf{X}}$ consisting of uncorrelated random numbers from a unit normal distribution is created. Then the corresponding $N$ by $1$ vector of correlated random variables, $\textit{\textbf{Y}}$, is evaluated by computing $\textit{\textbf{LX}}$ and subsequently adding the mean, $\boldsymbol{\mu}$ as
\begin{align}
    \textit{\textbf{Y}} = \textit{\textbf{L~X}} + \boldsymbol{\mu}.
\end{align}
Since the permeability has order of magnitude variations in a soil, the final transformation, $\textbf{k}=10^\textit{\textbf{Y}}$, yields the $N \times 1$ vector of absolute permeability inside each cell.  Then the porosity field, $\phi(\textbf{x})$, is then evaluated from the permeability by setting a maximum value of porosity, $\phi_{max}$ ($\phi_{max}$=85$\%$ in the present problem), and the maximum evaluated dimensionless permeability, $\Ki_{max}/\Ki_0$, from Equation \eqref{eq:constitutive-fxn} as $\phi(\textbf{x})=\phi_{max} \left(\Ki(\textbf{x})/\Ki_{max} \right)^{1/\m}$. Choosing a higher value of $\m$ ($\m=8$) helps fix the mean porosity to $46.5\%$ and minimum to $21.6\%$ which are close to the values for the fine sand \citep{das2008advanced}. The resulting porosity field is shown in Figure~\ref{fig:heterogeneous-porosity-field}.

\section{{Comparison against Hydrus}} \label{sec:hydrus1Dcomparison}
{In Section \ref{sec:test-problems} we have provided comparisons between the semi-analytic solutions and the numerical solutions of the governing equation in the limit of no capillary forces. Below we provide additional validation via a comparison with a numerical solution to the full Richards equation using Hydrus-1D \citep{vsimunek2012hydrus} and analytical results using kinematic wave theory \citep{shadab2022analysis}. It will help highlight the importance of neglecting capillary terms and dissect its effect at different spatial scales, along with solver performance.}

\subsection{{At smaller scales}} \label{sec:smaller-scale}
 Here we choose the infiltration in double-textured soil, as discussed in Section~\ref{sec:two-layer-test}. The Hydrus-1D simulation domain [$0,100$ cm] is discretized uniformly into 400 cells with cell width of $0.25$ cm and runs until a maximum time of 16.96 hours. The domain has a jump in porosity and hydraulic conductivity at $z_0=50$ cm (dashed line). Table~\ref{table:1} summarizes the properties utilized in Hydrus.

For the Hydrus simulations we use the modified van-Genuchten model (mvG), because the suction head derivatives $\d \Psi/\d s_w$ remain bounded when $s_w \to 1$, which is necessary for convergence in the presence of a saturated region \citep{Vogel1988}. The solver does not converge otherwise. The parameters for mvG model are also given in Table~\ref{table:1} and chosen so that upper and lower layers correspond to sandy-loam and silty clay loam \citep{carsel1988developing}, respectively. {The $Alpha$ and $n$ parameters of the mvG model \citep{Vogel1988} in the Hydrus software \citep{vsimunek2012hydrus} are chosen to reduce the capillary pressure effects to a minimum.} The upper (surface) boundary condition is atmospheric with surface runoff whereas the lower boundary condition is free drainage. The precipitation $R$ is $42.44$ cm/day. The initial head inside the domain is $-100$ cm. 

The need to use the mvG model in Hydrus leads to a discrepancy with our analytic model which is based on the Brooks-Corey model for hydraulic properties. However, due to the simplicity of the solution the hydraulic properties are only sampled at the saturation of the initial wetting front, in addition to the dry and fully saturated states. This allows us to choose the parameters in the Brooks-Corey model to match the hydraulic conductivity at the front saturation. The fitted Kozeny-Carman parameter is $m=3$ and Brooks-Corey parameter is $n=7.15306$ for the upper and lower layer porosities of $0.43$  and $0.1$ respectively.

\begin{table}[]
\caption{A summary of hydraulic properties of upper and lower layers for modified van-Genuchten model \protect\citep{Vogel1988} in Hydrus-1D \protect\citep{vsimunek2012hydrus}.}
\centering
\begin{tabular}{c c c c c c c c c c c}
\toprule
Layer&$Qr$ & $Qs$ & $Alpha$ (1/cm) & $n$ & $K_s$ (cm/day) & $l$& $Qm$ & $Qa$ & $Qk$ & $K_k$ (cm/day) \\
\midrule
Upper &0 & 0.43 & 0.5 & 2.68 & 106.1   & 0.5 &0.43 & 0 & 0.43 & 106.1 \\
Lower &0 & 0.1  & 0.5 & 2.68 & 1.33447 & 0.5 &0.1 & 0  & 0.1 & 1.33447\\
\bottomrule
\end{tabular}
\label{table:1}
\end{table}

The Hydrus simulation results are shown in Figure \ref{fig:12} together with the analytic solution from Section \ref{sec:two-layer-test} \citep{shadab2022analysis} and the numerical solution in the limit of no capillary forces from the proposed model. The agreement with the numerical solution is excellent in all three stages of the flow, but the fronts in the Hydrus solution are less sharp due to the presence of capillary forces which provide an additional diffusive water flux.

\begin{figure}
    \centering
    \includegraphics[width=0.8\linewidth]{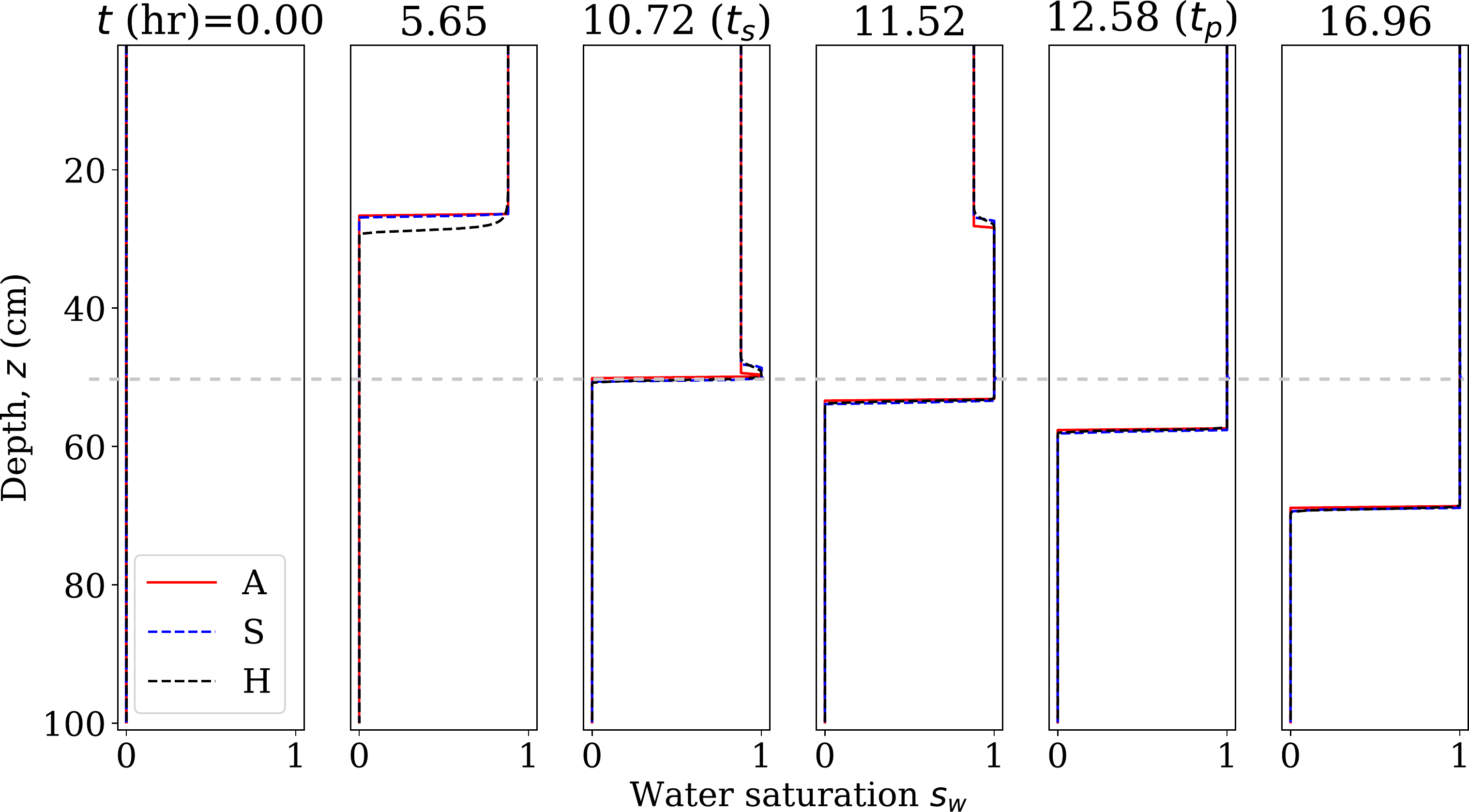}
    \caption{Infiltration process in an initially dry two-layered soil with $m=3,n=7.15306$ and no residual saturations shown at different times. The porosities $\phi_u=0.43$, $\phi_l=0.1$ and $R=42.44$ cm/day are fixed. The vertical axis refers to the depth $z$ whereas the horizontal axis refers to the water saturation $s_w$.  `A', `S', and `H' refer to the analytical, simulated (present model), and Hydrus solutions respectively. Here $t_s$ and $t_p$ correspond to the time of formation of the saturated region and time of ponding respectively.}
    \label{fig:12}
\end{figure}

\subsection{{At intermediate and larger scales}}\label{sec:large-scale}

\begin{figure}[!htbp]
    \centering
    \includegraphics[width=0.25\linewidth]{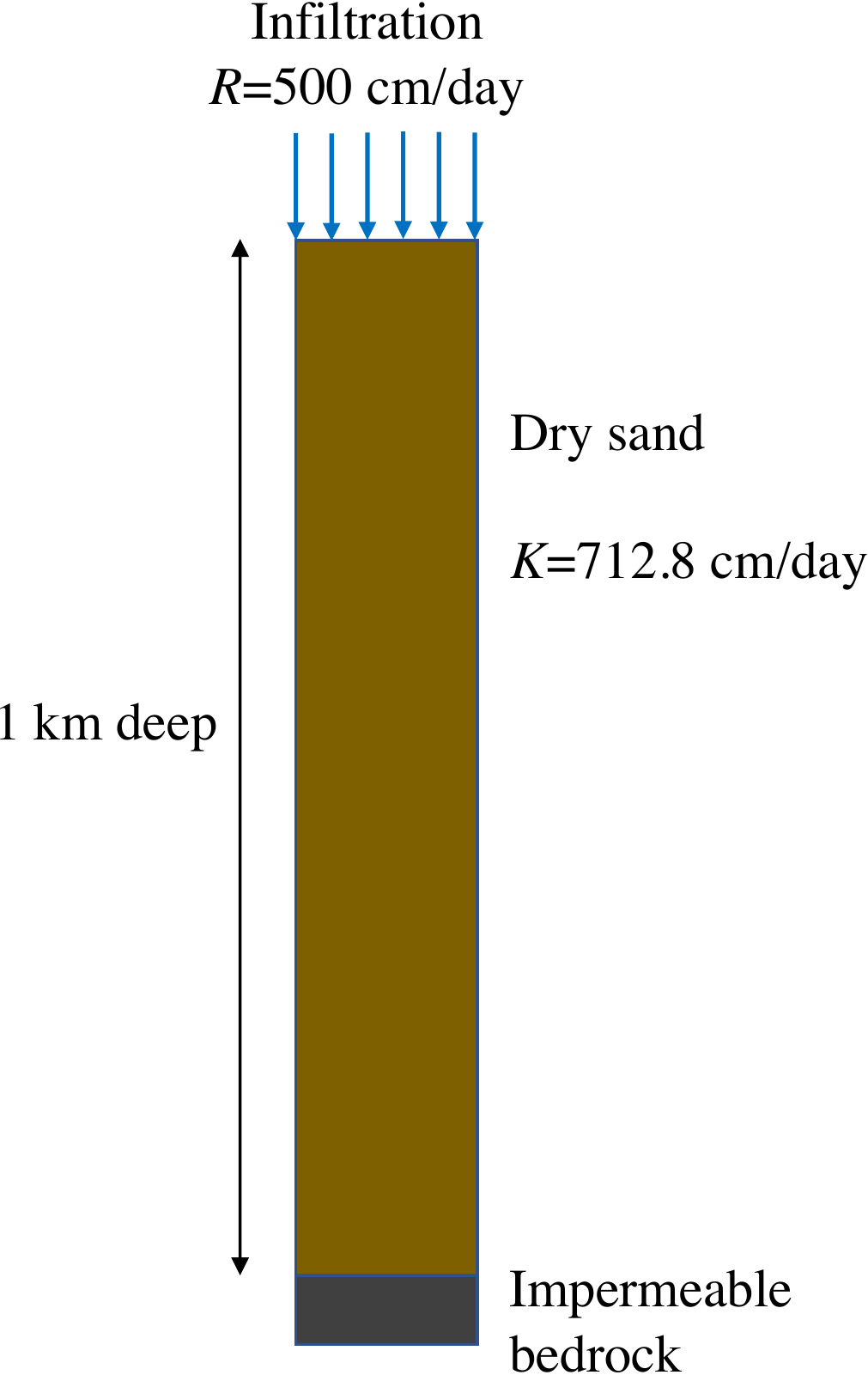}
    \caption{Schematic showing near-saturated, one-dimensional infiltration in dry sand at a large spatial scale.}
    \label{fig:representative-fig-hydrus}
\end{figure}

\begin{figure}[!htbp]
    \centering
    \includegraphics[width=0.5\linewidth]{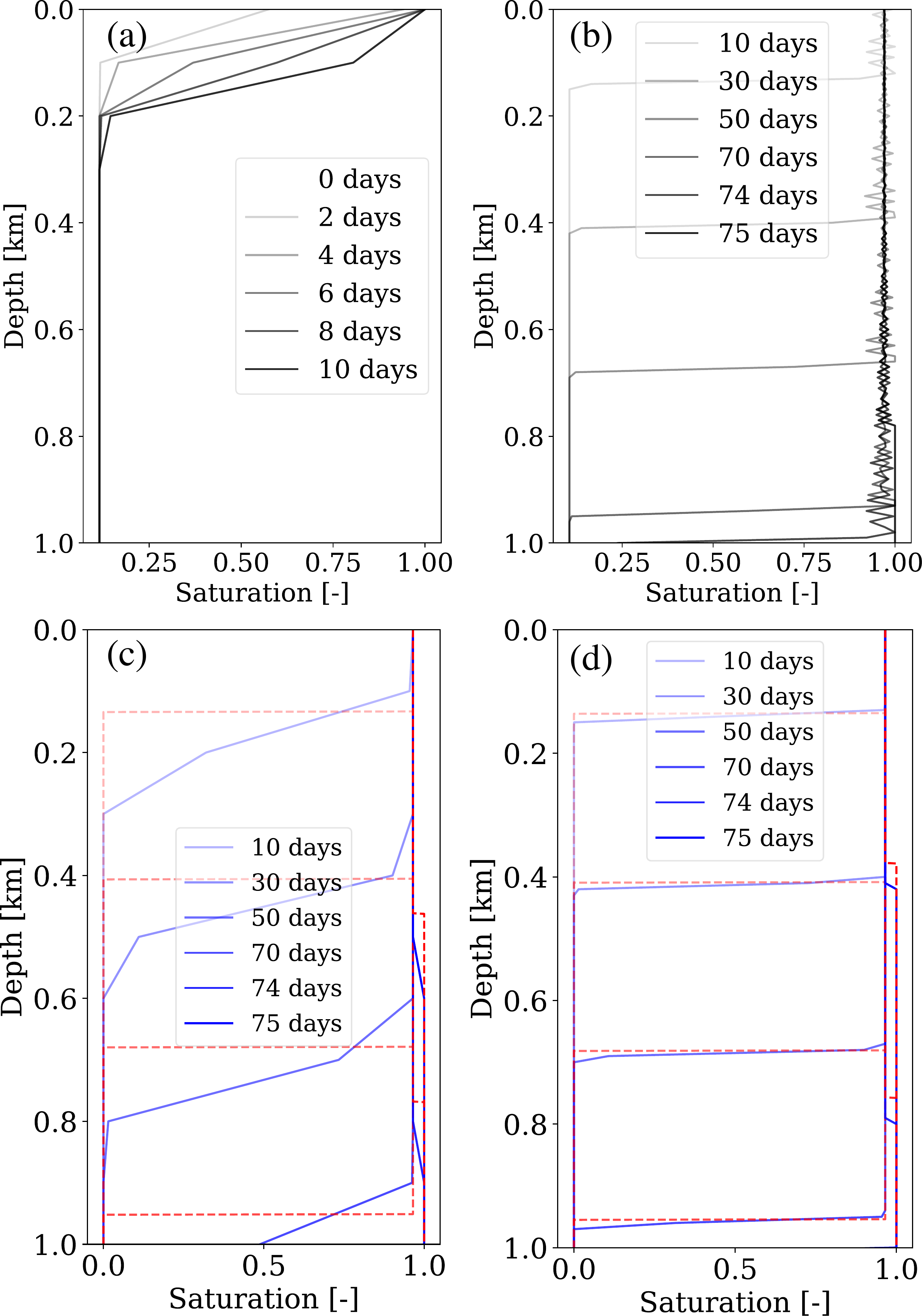}
    \caption{Saturation profiles at different times during large-scale 1D infiltration process in soil with impermeable base. Solutions are obtained from Hydrus-1D for (a) coarse grid (100 m cell size) and (b) fine grid (10 m cell size), and from the proposed method for (c) coarse grid (100 m cell size) and (d) fine grid (10 m cell size). The blue line shows calculated results from our solver whereas the red dashed line shows the corresponding analytic results from extended kinematic wave theory \citep{shadab2022analysis}. Panel \textit{a} is limited to 10 days as Hydrus blows up after it.}
    \label{fig:infiltration-figures}
\end{figure}

\begin{figure}[!htbp]
    \centering
    \includegraphics[width=0.65\linewidth]{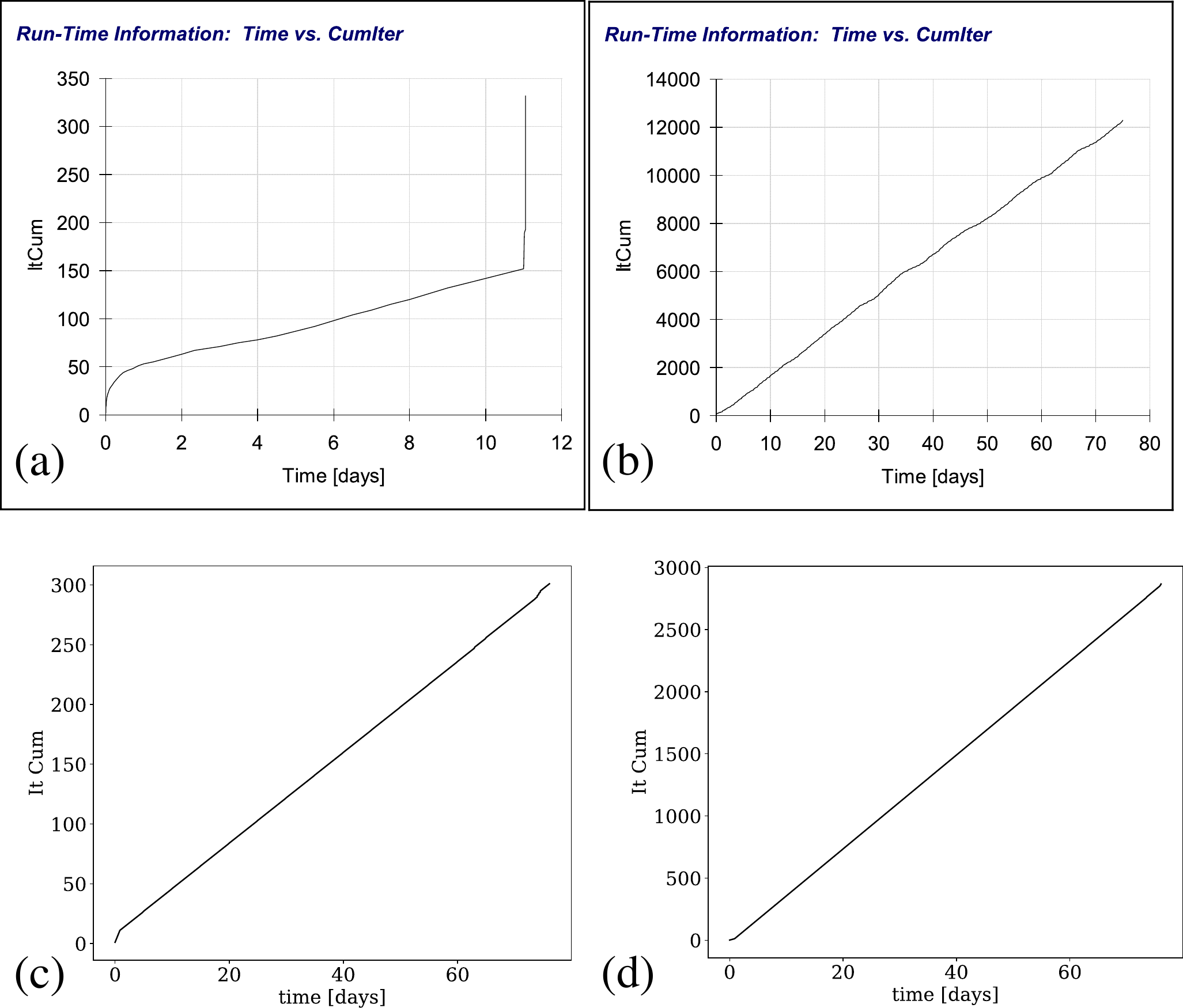}
    \caption{Cumulative iterations for the tests performed corresponding to each panel in Figure \ref{fig:infiltration-figures}. Therefore, the curves are obtained from Hydrus for (a) coarse grid (100 m cell size) and (b) fine grid (10 m cell size), and from the proposed method for (c) coarse grid (100 m cell size) and (d) fine grid (10 m cell size).}
    \label{fig:cum-iter}
\end{figure}

\begin{figure}[!htbp]
    \centering
    \includegraphics[width=0.5\linewidth]{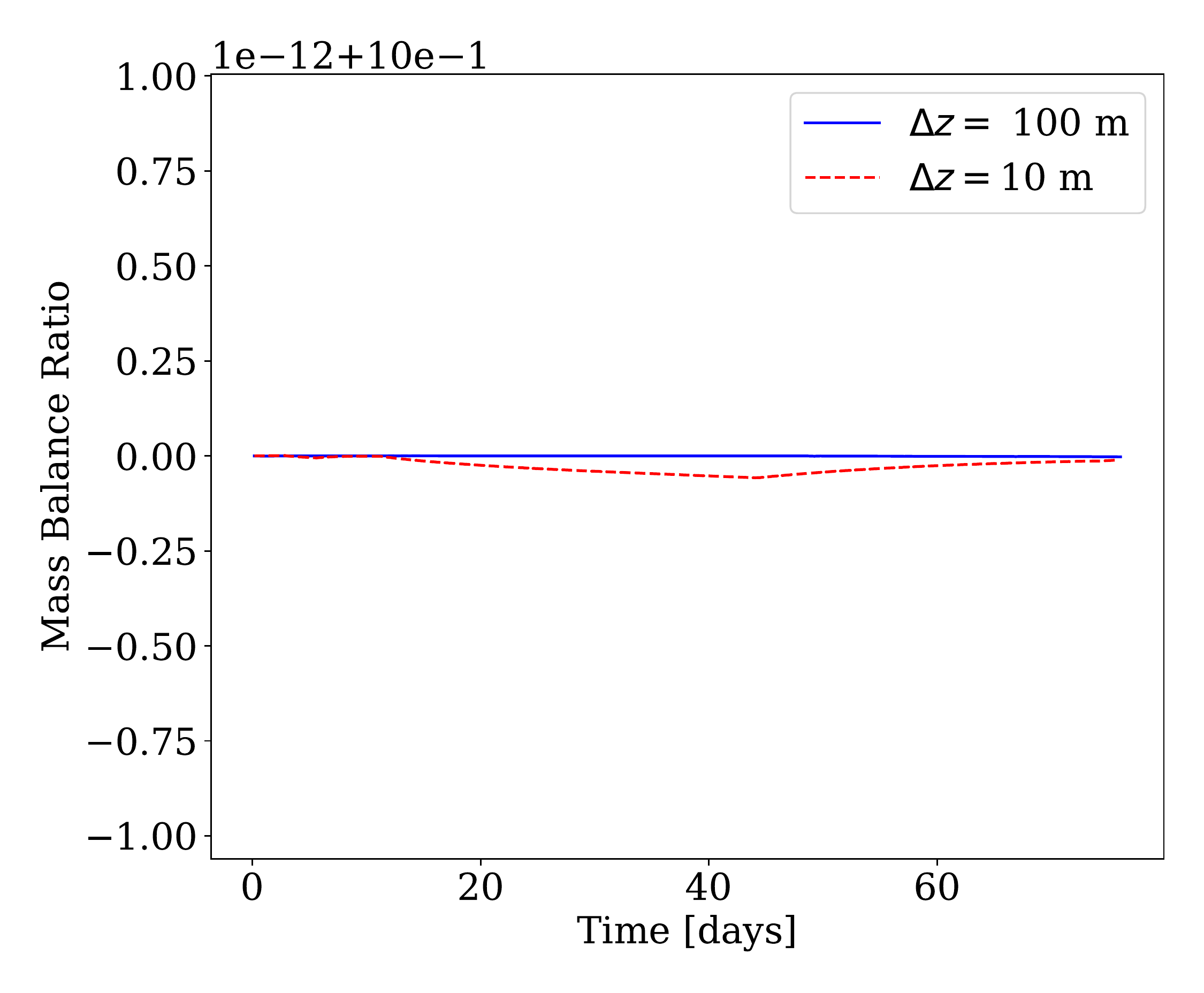}
    \caption{Mass balance ratio for the test shown in Figure \ref{fig:cum-iter} on coarse (100 m cell size) and fine grids (10 m cell size) using the proposed numerical method.}
    \label{fig:Massbalance}
\end{figure}

Next, we consider a simple test involving near-saturated infiltration in a homogeneous 1D column of dry sand over impermeable bedrock (Figure \ref{fig:representative-fig-hydrus}). This problem is studied Hydrus 1D (for sand), our proposed model and analytic solutions. We utilize the standard properties of sand provided in Hydrus software ($Qr=0,Qs=0.43,K_s=712.8$ cm/day, $n=2.68$). The domain of depth $z \in$ [0,1 km] is divided uniformly into 10 and 100 cells with two different cell sizes (100 m and 10 m), referred to as coarse and fine resolution below. The column chosen is large to illustrate the practical challenges arising in modeling large systems where the capillary transition may not be resolved by the grid. The resulting solution is a wetting front that moves downwards followed by a rapidly rising perched water table that forms around 73 days due to ponding on the impermeable bedrock at the base. The analytic result again comes from the kinematic wave theory for double-textured soil, as given in Section \ref{sec:two-layer-test} \citep{shadab2022analysis}.

We found that at coarse resolution, Hydrus 1D fails to converge after 10 days possibly due to capillary pressure term (Figure \ref{fig:infiltration-figures}\emph{a}) whereas our solver converges without problems and captures the front speed, though numerical diffusion broadens the front compared to the analytic solution (red-dashed) (Figure \ref{fig:infiltration-figures}\emph{c}). Increasing the grid resolution to 10 m, allows Hydrus to converge, but the solution is highly oscillatory near the wetting front (Figure \ref{fig:infiltration-figures}\emph{b}). At the same grid resolution, our code (Figure \ref{fig:infiltration-figures}\emph{d}) provides a sharp wetting front without oscillations and compares well with the analytic solution. In addition we noticed that the Hydrus solution does not conserve mass (not shown). 

This test problem also allows a simple comparison of the numerical efficiency of both methods. Figure \ref{fig:cum-iter} compares the cumulative iterations of each simulator on both the fine and the coarse grid. On the coarse grid, Figure \ref{fig:cum-iter}\textit{a} illustrates the failure of Hydrus to converge at approximately 11 days. Just before the solution diverges, Hydrus has used approximately 150 iterations while our code shown in Figure \ref{fig:cum-iter}\textit{c} has used only approximately 60 iterations. On the fine grid both simulators converge and have a linear increase in cumulative iterations with time. The total number of iterations required by Hydrus is  14,000 (Figure \ref{fig:cum-iter}\textit{b}) while our code just requires 3000 iterations (Figure \ref{fig:cum-iter}\textit{d}). 

{Lastly, the conservation of mass is studied using the mass balance ratio criteria defined by \cite{Celia1990} as:
\begin{equation}
\textrm{Mass balance ratio} = \frac{\textrm{Total additional mass in the domain}}{\textrm{Total net flux of water into the domain}}.
\end{equation}
A unity mass balance ratio signifies perfect conservation of mass for a numerical method. The proposed numerical technique utilizes the conservation form of the governing equation and conserves mass for both tests at coarse and fine grids (Figure~\ref{fig:Massbalance}). The drop in the mass balance ratio is on the order of $10^{-12}$ due to machine precision. For conventional numerical methods, it may significantly drop by an order of $10^{-1}$ \citep[see][]{Celia1990} showing a lack of discrete mass conservation.}

\printcredits

\section*{Code availability}
The proposed model is referred to as \textit{VarSatFlow} (Variably Saturated groundwater Flow). All codes to generate figures in the manuscript are available on GitHub: \url{https://github.com/mashadab/VarSatFlow} \citep{shadab2024varsatflow}.

\bibliographystyle{cas-model2-names}

\bibliography{marc,cas-refs}





\end{document}